\documentclass[conference]{IEEEtran}

\newif\ifarxiv\arxivtrue

\IEEEoverridecommandlockouts

\usepackage[LGR,T1]{fontenc}
\usepackage{graphicx}
\usepackage{proof}
\usepackage{amsmath}
\usepackage{amssymb}
\usepackage{amsthm}
\usepackage{stmaryrd}
\usepackage{tensor}
\usepackage[all]{xy}
\usepackage{bbm}
\usepackage{comment}
\usepackage{xspace}
\usepackage{tikz-cd}
\usepackage{mathrsfs}
\usepackage{subcaption}
\usepackage{algorithm}
\usepackage{xcolor}
\usepackage{enumerate}
\usepackage{listings}
\usepackage{nicefrac}
\lstdefinestyle{mystyle}{
    language=ML,
    commentstyle=\color{green},
    keywordstyle=\color{magenta},
    keywordstyle = [2]{\color{blue}},
    basicstyle=\ttfamily\small,
    breakatwhitespace=false,         
    breaklines=true,                 
    captionpos=b,                    
    keepspaces=true,                                
    numbersep=5pt,                  
    showspaces=false,                
    showstringspaces=false,
    showtabs=false,                  
    tabsize=2,
    morekeywords={and,or,not},
    morekeywords=[2]{fst,snd,map,reweight,prng,tl,hd,cast},
    mathescape=true,
    frame=single
}
\usepackage{cleveref}

\crefname{figure}{Fig.}{Figs.}
\crefname{table}{Table}{Tables}
\crefname{section}{\S}{\S}
\crefname{subsection}{\S}{\S}
\crefname{proposition}{Prop.}{Props.}
\crefname{theorem}{Thm.}{Thms.}
\crefname{example}{Ex.}{Exs.}
\crefname{listing}{Listing}{Listings}
\crefname{corollary}{Cor.}{Cors.}
\crefname{equation}{Eq.}{Eqs.}
\crefname{appendix}{Appendix}{Appendices}

\newenvironment{appendixproof}[1][Proof]
    {\noindent \textbf{Proof of #1.}
    \\
    }
    {\qed
	\\    
    }
\newtheorem{theorem}{Theorem}[section]

\newtheorem{proposition}{Proposition}[section]
\newtheorem{example}{Example}[section]
\newtheorem{definition}{Definition}[section]

\DeclareSymbolFont{ttgreek}{LGR}{cmtt}{m}{n}

\newcommand{\resp}{resp.\@\xspace}

\newcommand{\eg}{e.g.\@\xspace}
\newcommand{\ie}{i.e.\@\xspace}
\newcommand{\io}{i.o.\@\xspace}
\newcommand{\wrt}{w.r.t.\@\xspace}

\newcommand{\aka}{a.k.a.\@\xspace}

\newcommand{\defeq}{\triangleq}
\newcommand{\R}{\mathbb{R}}

\newcommand{\Rnn}{\mathbb{R}_{\geq 0}}
\newcommand{\N}{\mathbb{N}}
\newcommand{\B}{\mathbb{B}}

\newcommand{\inv}{^{-1}}

\newcommand{\pr}{\mathcal{P}}
\newcommand{\prb}{\pr_\bot}

\newcommand{\pnt}[1][T]{p^N_{\Code{#1}}}
\newcommand{\predinv}[2]{{#1}\inv(#2)}

\newcommand{\denote}[1]{\left\llbracket #1\right\rrbracket}

\newcommand{\Code}[1]{\mathtt{#1}}
\DeclareMathSymbol{\stypel}{\mathord}{ttgreek}{`S}
\newcommand{\stype}{\stypel\,}
\newcommand{\tl}{\Code{tl}}
\newcommand{\hd}{\Code{hd}}
\newcommand{\map}{\Code{map}}
\newcommand{\fst}{\Code{fst}}
\newcommand{\snd}{\Code{snd}}
\newcommand{\rew}{\Code{reweight}}
\newcommand{\wt}{\Code{wt}}
\newcommand{\prng}{\Code{prng}}

\newcommand{\la}[3]{\lambda #1:#2~.~#3}

\newcommand{\ite}[3]{\Code{if}~#1~\Code{then}~#2~\Code{else}~#3}
\newcommand{\letx}[3]{\Code{let}~#1=#2~\Code{in}~#3}
\newcommand{\sub}{\triangleleft}
\newcommand{\cast}[2]{\Code{cast}\langle \Code{#1}\rangle #2}
\newcommand{\case}[5]{\Code{case}~#1~\Code{of}~\left\{\left(#5,#2_{#5}\right)\Rightarrow #3_{#5}\right\}_{#5\in #4}}
\newcommand{\ini}[2]{\Code{in}_{#1}\left(#2\right)}
\newcommand{\pb}[3]{\tensor*[_{#1}]{\Code{#3}}{_{#2}}}

\begin{document}

\title{Deterministic stream-sampling for probabilistic programming: semantics and verification
\thanks{This work was supported by the Leverhulme Project Grant “Verification of Machine Learning Algorithms”.}
}

\author{\IEEEauthorblockN{Fredrik Dahlqvist}
\IEEEauthorblockA{\textit{Queen Mary University of London} \\
and \textit{University College London}\\
London,  United Kingdom \\
f.dahlqvist@qmul.ac.uk}
\and
\IEEEauthorblockN{Alexandra Silva}
\IEEEauthorblockA{\textit{Department of Computer Science} \\
\textit{Cornell University}\\
Ithaca,  USA \\
alexandra.silva@cornell.edu}
\and
\IEEEauthorblockN{William Smith}
\IEEEauthorblockA{\textit{Department of Computer Science} \\
\textit{University College London}\\
London, United Kingdom  \\
william.smith.19@ucl.ac.uk}
}

\maketitle

\begin{abstract} 
Probabilistic programming languages rely fundamentally on some notion of sampling, and this is doubly true for probabilistic programming languages which perform Bayesian inference using Monte Carlo techniques.
Verifying samplers---proving that they generate samples from the correct distribution---is crucial to the use of probabilistic programming languages for statistical modelling and inference.
However, the typical denotational semantics of probabilistic programs is incompatible with deterministic notions of sampling. This is problematic, considering that most statistical inference is performed using pseudorandom number generators.

We present a higher-order probabilistic programming language centred on the notion of  \emph{samplers} and \emph{sampler operations}. We give this language an operational and denotational semantics in terms of continuous maps between topological spaces. Our language also supports discontinuous operations, such as comparisons between reals, by using the type system to track discontinuities. This feature might be of independent interest, for example in the context of differentiable programming.

Using this language, we develop tools for the formal verification of sampler correctness. We present an equational calculus to reason about equivalence of samplers, and a sound calculus to prove semantic correctness of samplers, \ie that a sampler correctly targets a given measure by construction.
\end{abstract}


\section{Introduction}

Probabilistic programming languages without conditioning -- that is to say, programming languages capable of drawing random samples -- and the concepts of Monte Carlo methods and randomized algorithms have been around as long as true computers have\footnote{See \cite{blackwell2019usability} for an overview of probabilistic programming languages, \cite{brooks2011handbook} for a historical overview of the Monte Carlo method, and \cite{historyuniform2017} for a history of pseudorandom number generation.}; however, the introduction of languages with conditioning, higher-order features, continuous variables, recursion, and their application to statistical modelling and machine learning, is a product of the twenty-first century \cite{thomas1994bugs,winbugs2000,goodman2012church,venture2014,wood-aistats-2014,carpenter2017stan,bingham2019pyro}. Since a probabilistic programming language with conditioning must come equipped with a range of inference algorithms and sampling methods, and since the rate of introduction of these has increased in recent years,  new formal methods must be developed for the \emph{verification} of these algorithms.

We aim to make the verification of inference algorithms straightforward by introducing a language endowed with a sampler type, featuring many of the sampler operations used in these inference algorithms, and a calculus for reasoning about correctness of these samplers relative to intended `target' distributions. The semantics of our language is fully deterministic, in order to allow the use of deterministic -- pseudorandom -- samplers, in a simple manner and without paradox.


When assigning operational and denotational semantics to probabilistic programs, an interesting asymmetry emerges: the simplest reasonable denotational semantics of a first-order language with continuous datatypes is in terms of probability measures \cite{K81c}, while the simplest reasonable operational semantics is in terms of sampled \emph{values}  -- the latter being much closer to the intuitions used by programmers of languages with the ability to draw samples.

The denotational semantics of probabilistic programming languages, in terms of measures (broadly construed) is well-understood \cite{K81c,heunen2017convenient,ehrhard2017measurable,vakar2019domain,dahlqvist2019semantics}. The most common approaches to the operational semantics of such a language, as highlighted by \cite{ehrhard2017measurable}, are \emph{trace semantics} and \emph{Markov chain semantics}. The latter is the chosen operational semantics for a number of probabilistic $\lambda$-calculi \cite{ehrhard2011computational,lago2012probabilistic,borgstrom2016lambda,ehrhard2018full,ehrhard2017measurable,faggian2019lambda}  and probabilistic languages \cite{staton2016semantics,vakar2019domain},  but does not speak of sampled values, only of distributions on execution paths. From the perspective of the asymmetry described above, it is thus closer to a denotational semantics.

Trace semantics, originally developed in \cite{K81c} and later applied in \cite{park2005probabilistic,borgstrom2016lambda,dahlqvist_silva_kozen_2020,bagnall2023formally}, assumes that for each distribution in the language, an infinite set of samples has been produced ahead-of-time.  When a sample is requested, the head of this sequence is popped and used in the computation, and the tail of the sequence is kept available for further sampling. This perspective models samplers, such as $\Code{rand()}$, as functions with hidden side effects on the state of the machine -- in line with a programmer's intuition on the nature of sequential calls to $\Code{rand()}$. The natural notion of adequacy with respect to the denotational semantics is to show that subject to the assumption that each element of these sequences is sampled independently from its corresponding distribution, the resulting pushforward through the program is identical to the program's denotational semantics. We see two issues with this approach.

First, the supposition that all samples are pre-computed ahead-of-time is incompatible with pseudorandom generation of `random' values, since computationally-generated samples and truly-random samples are in fact distinguishable. For example, if \((x_0, x_1, \ldots)\) is a sequence of samples targeting the distribution \(P\) which was produced via iteration of a computable map \(x_{n+1} = T(x_n)\), then the program \(T(x) - x\) will behave differently if \(x\) is a `truly random' sample from \(P\) than if \(x\) is produced by the aforementioned iterative procedure, and so the operational and denotational semantics no longer cohere; similar counterexamples exist for any pseudorandom number generator.
If \((x_0, x_1, \ldots)\) is a deterministic sequence, meaningful coherence between the operational and denotational sequence can only be assured if it is assumed that this sequence is \emph{Martin-L\"of random} (first defined in \cite{martin1966definition}, later generalised to computable metric spaces in \cite{hoyrup2009computability}); unfortunately, all Martin-L\"of random sequences are uncomputable. Pseudorandom numbers are in fact used in simulation far more commonly than `true' physical randomness, as they are typically faster to obtain, have the advantage of being reproducible given a particular seed, and, subject to certain assumptions, can even have better convergence properties\ifarxiv\footnote{This remark is in reference to the related field of quasi-Monte Carlo techniques, outlined in \cite{leobacher2014quasimc}.}.\else\cite{leobacher2014quasimc}.\fi
We find the inability of trace semantics to describe pseudorandom number generation to be a significant weakness.

Second, from the trace semantics perspective, the notion of a sampler type is inextricably bound up in issues regarding side-effects, which makes verification challenging. In order to properly assign the syntax $\Code{rand}$, without parentheses, a meaningful semantics as a function, we must give it a monadic interpretation as in \cite{park2005probabilistic}, accounting for its hidden effect on the trace. The correctness of code which inputs and outputs samplers -- sampler operations -- is then subject to the state of the trace when computation is started, which is not contained within either the code of the sampler operation in question or the code of the samplers it inputs. This pattern, of using sampler operations to create composite samplers which make use of other `primitive' samplers, is a central theme in computational statistics. In particular, inference algorithms within Bayesian statistics, to which probabilistic programming languages with conditioning compile, make heavy use of this technique. Commonly-used sampling methods, such as importance sampling, rejection sampling, and particle Markov chain Monte Carlo methods, are naturally understood as composite samplers of this type \cite{doucet2010,robert2013monte}. We prefer a side-effect-free perspective from which the correctness of sampler operations can be demonstrated subject to assumptions about the samplers input to these programs, as opposed to a perspective in which their correctness depends also on the state of the machine on which these operations are run.



\noindent{\bf Contributions.} We develop a language based around the idea of sampler types and sampler operations, which allows reasoning about deterministic and real-valued samplers, and which is designed to make verification of these samplers natural. A syntax, operational semantics, and denotational semantics for our language are introduced in \cref{sec:language}, and an adequacy result relating them is shown. \cref{sec:equivalence} lays out a notion of equivalence of samplers, which will be applied to simplify programs. Finally, in \cref{sec:verification}, we discuss methods for proving that samplers target the desired probability measure (i.e. verifying samplers), and introduce a sound calculus for verifying the correctness of composite samplers which is capable of demonstrating the soundness of common Monte Carlo techniques such as importance sampling and rejection sampling.

\section{Examples}\label{sec:examples}

The purpose of this section is twofold. First, we present examples of how samplers are transformed in order to create new samplers. We use this opportunity to informally introduce a language with a sampler type constructor $\stype$ and operations for constructing and manipulating samplers. 
Second, we present techniques to reason about the correctness of sampling algorithms.  These come in two flavours. We reason \emph{equationally} about the equivalence between samplers (see \cref{sec:equivalence}),  and we reason \emph{semantically} about whether a sampler does the job it is designed to do -- namely, generate deviates from a target distribution (see \cref{sec:verification}).

\subsection{Von Neumann extractor}\label{sec:examplevonneumann}

\begin{figure*}[t]
\centering
\footnotesize
\[
\infer{\Code{\vdash map(proj, reweight(choice, flip^2)) : \stype B} \rightsquigarrow \denote{\Code{proj}}_* (\denote{\Code{choice}} \cdot \text{Ber}(p)^2) = \text{Ber}(\nicefrac{1}{2})}
{\infer
{\Code{\vdash reweight(choice, flip^2) : \stype (B \times B)} \rightsquigarrow \denote{\Code{choice}} \cdot \text{Ber}(p)^2}
{\Code{\vdash flip^2 : \stype (B \times B)} \rightsquigarrow \text{Ber}(p)^2 & \Code{\vdash choice : B \times B \to R^+}} & \Code{\vdash proj : B \times B \to B}}
\]
\vspace{-4mm}
\caption{Validity of von Neumann extractor}
\vspace{-4mm}
\label{fig:extractorproof}
\end{figure*}

\begin{figure*}[t]
{
\footnotesize
\centering
\[
\infer{\Code{\vdash reweight(phi, map(plus, rand^2)) : \stype R} \rightsquigarrow \denote{\Code{phi}} \cdot (\denote{\Code{plus}}_* U^2) = P}
{\infer{\Code{ \vdash map(plus, rand^2) : \stype R} \rightsquigarrow \denote{\Code{plus}}_* U^2}
{\Code{\vdash rand^2 : \stype (R \times R)} \rightsquigarrow U^2 & \vdash \Code{plus : R \times R \to R}} & \Code{\vdash phi : R \to R^+}}
\]
}
\vspace{-4mm}
\caption{Validity of importance sampling in \cref{alg:importancesampling}}
\vspace{-4mm}
\label{fig:importancesamplingproof}
\end{figure*}

We begin with a simple family of discrete samplers known as von Neumann extractors. This example will illustrate the concept of a sampler's \emph{self-product}, a central concept for a language featuring sampler types.
The von Neumann extractor~\cite{trevisan2000extracting} is a simple procedure which, given flips from a \emph{biased} coin on \(\{\Code{True}, \Code{False}\}\) with probability \(p \in (0, 1)\) of landing \(\Code{True}\), produces flips from an \emph{unbiased} coin with probability \(\nicefrac{1}{2}\) of landing \(\Code{True}\). We view this as a sampler \(\Code{v}\) of Boolean type -- notation $\Code{v}:\stype\Code{B}$ -- which, given another Boolean-valued sampler \(\Code{flip : \stype B}\) representing our biased coin, constructs an unbiased Boolean-valued sampler.
A simple implementation of the von Neumann extractor is given in \cref{alg:vonneumannextractor}. 

\lstset{style=mystyle}

\begin{lstlisting}[caption=von Neumann extractor, label={alg:vonneumannextractor}]
	let choice = $\lambda$b : B $\times$ B .
	if (fst(b) and snd(b)) or (not fst(b) and not snd(b)) then 0 else 1
	in let proj = $\lambda$b : B $\times$ B . fst(b)
	in map(proj, reweight(choice, flip$^2$))
\end{lstlisting}

The idea behind this algorithm is that if \((b_1, b_2)\) are sampled independently from a Bernoulli distribution with parameter \(p\), then the probabilities of the outcomes \((b_1, b_2) = (\Code{False}, \Code{True})\) and \((b_1, b_2) = (\Code{True}, \Code{False})\) are both \(p(1-p)\), and so the first element \(b_1\) of each pair is an unbiased flip. In \cref{alg:vonneumannextractor}, samples \((b_1, b_2)\) where \(b_1 = b_2\) are removed by the \(\Code{reweight}\) operation, which sets the \emph{weight} of such pairs to zero; then, the command \(\Code{map}\) applies the function \(\Code{proj}\) to each pair \((b_1, b_2)\), returning the sampler whose outputs are only the first element \(b_1\).

Note the appearance of \(\Code{flip}^2\) in the von Neumann extractor; this is the `self-product' of the sampler \(\Code{flip}\). Where \(\Code{flip : \stype B}\) produces Boolean-valued samples, \(\Code{flip^2 : \stype(B \times B)}\) produces samples which are pairs of Booleans. Given a sampler \(t:\stype \Code{T}\) of type \(\Code{T}\), the self-product \(t^2 : \stype (\Code{T} \times \Code{T})\) is a sampler whose elements are adjacent samples from \(\Code{T}\); the same construction, detailed in~\cref{sec:language}, is easily extended to arbitrary self-powers \(t^K : \stype(\Code{T}^K)\).


The main application of our language is to serve as a setting for the formal verification of its samplers. Let \(\Code{v : \stype B}\) be the von Neumann extractor defined in \cref{alg:vonneumannextractor}; the task of verifying \(\Code{v}\) is the task of showing that \(\Code{v}\) `targets' the uniform distribution \(\text{Ber}\left(\nicefrac{1}{2}\right)\) -- meaning, informally, that in the limit of increasing sample size, \(\Code{v}\) generates unbiased flips. In \cref{sec:asymptotictargeting}, we define a relation \(\rightsquigarrow\) between samplers and distributions which reifies this notion: we read \(\vdash \Code{v : \stype B} \rightsquigarrow \text{Ber}\left(\nicefrac{1}{2}\right)\) as `the sampler \(\Code{v}\) targets the measure \(\text{Ber}\left(\nicefrac{1}{2}\right)\)'.
The aforementioned self-product operation plays a crucial role in sampler verification, as it does \emph{not} suffice, in order to conclude that \(\Code{v}\) targets \(\text{Ber}\left(\nicefrac{1}{2}\right)\), to assume that \(\Code{flip}\) targets \(\text{Ber}(p)\) for some \(p \in (0, 1)\). Instead, we are required to make the stronger assumption that \(\Code{flip}^2\) targets \(\text{Ber}(p)^2\): in essence, that adjacent samples from \(\Code{flip}\) act as if they are independent. Following the literature on pseudorandom number generation, we refer to this property as \emph{\(K\)-equidistribution} (in this case, for \(K=2\)): we will say that a sampler \(s\) is \(K\)-equidistributed with respect to the distribution \(P\) if \(s^K\) targets \(P^K\). For example,  the assertion that a pseudorandom number generator which takes values in \(\{0, \ldots, N-1\}\) is \(K\)-equidistributed with respect to the uniform distribution is the assertion that all \(K\)-length words \(w \in \{0, \ldots, N-1\}^K\) are produced in equal proportion. Several commonly-used discrete PRNGs, such as xorshift and the Mersenne twister, have well-known \(K\)-equidistribution guarantees; see \cite{vigna2017marsaglia,matsumoto1998mersenne}.

In our calculus for asymptotic targeting, the validity of the von Neumann extractor is shown in \cref{fig:extractorproof}. Before we begin this derivation, it must be shown that the von Neumann extractor \(\Code{v}\) is equivalent to the simplified sampler \(\Code{map(proj, reweight(choice, flip^2))}\) in a context in which access to a Boolean-typed sampler \(\Code{flip}\) is assumed, where \(\Code{proj}\) and \(\Code{choice}\) are defined as in \cref{alg:vonneumannextractor}. We write this equivalence as \(\Code{flip : \stype B} \vdash \Code{v \approx map(proj, reweight(choice, flip^2)) : \stype B}\); this equivalence relation is discussed in \cref{sec:equivalence}, and is shown in this particular case using the let-binding rule of \cref{table:equivalencerules}. 

Having rewritten $\Code{v}$ in this way, and using the hypothesis of 2-equidistribution \(\Code{\vdash flip^2 : \stype B} \rightsquigarrow \text{Ber}(p)^2\), we derive our conclusion in \cref{fig:extractorproof} by applying the rules from \cref{sec:asymptotictargeting} corresponding to the sampler operations \(\Code{reweight}\) and \(\Code{map}\). These rules show us that \(\Code{v}\) targets the measure \(\denote{\Code{proj}}_* (\denote{\Code{choice}} \cdot \text{Ber}(p)^2)\). (Here, as we will discuss in \cref{sec:asymptotictargeting}, \(\denote{f}_* \mu\) denotes the pushforward of the measure \(\mu\) through the function \(f\), and the notation \(f \cdot \mu\) denotes the measure \(\mu\) reweighted by the density \(f\).) To complete the proof, we show that this measure is identical to \(\text{Ber}(\nicefrac{1}{2})\), the uniform measure on \(\Code{B}\); this is straightforward. It is easily seen first that \(\denote{\Code{choice}} \cdot \text{Ber}(p)^2\) assigns probability \(\nicefrac{1}{2}\) to the samples \((\Code{True}, \Code{False})\) and \((\Code{False}, \Code{True})\) and zero probability to all other samples; the desired result then follows by observing that the function \(\Code{proj}\) simply drops the second sample.



\subsection{Importance sampling}\label{sec:exampleimportancesampling}
\begin{figure*}[t]
{
\footnotesize
\centering
\[
\infer{\vdash\Code{map(proj, reweight(accept, tri \otimes rand)) : \stype R} \rightsquigarrow \denote{\Code{proj}}_* (\denote{\Code{accept}} \cdot (\text{Tri} \otimes U)) = P}
{\infer
{ \vdash \Code{reweight(accept, tri \otimes rand) : \stype T} \rightsquigarrow \denote{\Code{accept}} \cdot (\text{Tri} \otimes U)}
{\Code{\vdash tri \otimes rand : \stype T} \rightsquigarrow \text{Tri} \otimes U & \vdash \Code{accept : T \to R^+}} & \Code{\vdash proj : T \to R}
}
\]
}
\vspace{-4mm}
\caption{Validity of rejection sampling in \cref{alg:rejectionsampling}}
\vspace{-4mm}
\label{fig:rejectionsamplingproof}
\end{figure*}

A central application of the reweighting operation is its role in \emph{importance sampling}. This is a commonly used technique \cite{robert2013monte,brooks2011handbook} in Bayesian learning and statistical inference, which transforms samples from a `proposal' distribution \(Q\) on the latent space \(X\) into approximate samples from a `target' distribution \(P\), where \(P\) is absolutely continuous with respect to \(Q\) with Radon-Nikodym derivative \(\frac{dP}{dQ}(x)\). Its operation is straightforward: for each sample \(x_n \sim Q\), compute the sample's weight \(w_n = \frac{dP}{dQ}(x)\), and then the \emph{weighted sample} \((x_n, w_n)\) is informally understood as an approximate sample from the target \(P\). Formally, the normalised empirical measure \(\sum_{n=1}^N \frac{w_n}{\sum_{i=1}^N w_i} \delta_{x_n}\), where \(\delta_x\) is the Dirac measure at \(x \in X\), converges weakly as \(N \to \infty\) to the target measure \(P\).

For example, consider the Bayesian inference problem in which the prior \(P_0\) is the triangular distribution on \([0, 2]\) and the likelihood of the observation \(y=3\) given the latent value \(x\) is a standard Gaussian \(L(x) = \frac{1}{\sqrt{2\pi}} \exp\left(-\frac{(3-x)^2}{2}\right)\). Let \(P\) represent the corresponding posterior distribution, whose density is proportional to the pointwise product of the triangular and Gaussian densities; a simple importance-sampling procedure for sampling from \(P\) in our language is shown in \cref{alg:importancesampling}. Here, we assume access to a sampler \(\Code{rand}\) which targets the uniform distribution on \([0, 1]\); recalling that a triangular random variable is the sum of two independent uniform random variables, and assuming \(\Code{rand}\) has the necessary independence property of 2-equidistribution, we sum two draws from \(\Code{rand}\) to produce a triangular random variable. Finally, we \emph{reweight} the result according to the likelihood \(L(x)\), yielding a sampler which targets the posterior distribution \(P\) corresponding to the observed datum \(y=3\).


\begin{lstlisting}[caption=Importance sampling, label={alg:importancesampling}]
let phi = $\lambda$x : R . 1/sqrt(2*pi) * exp(-1/2*(3-x)*(3-x))
	in let plus = $\lambda$u : R $\times$ R . fst(u) + snd(u)
		in reweight(phi, map(plus, rand$^2$))
\end{lstlisting}



The validity of this sampler -- \ie the fact that it targets the correct posterior distribution -- follows easily in our targeting calculus. Under the hypothesis that \(\Code{rand}\) produces 2-equidistributed samples with respect to the uniform distribution \(U\), the derivation \cref{fig:importancesamplingproof} proves that the sampler defined in \cref{alg:importancesampling} targets the measure \(\denote{\Code{phi}} \cdot (\denote{\Code{plus}}_* U^2)\). It remains to show that this measure is the desired \(P\); once one shows that the sum of two independent uniform variates is triangular, this follows by definition of the reweighting operation \(\cdot\). The same argument suffices for any observation \(y\).

\subsection{Rejection sampling}\label{sec:examplerejectionsampling}

Our final example of sampler verification is an instance of the technique known as \emph{rejection sampling}. \cref{alg:rejectionsampling} applies rejection sampling from the prior to the same Bayesian inference problem discussed in \cref{sec:exampleimportancesampling} to yield a sampler which targets the same posterior distribution \(P\). Of particular importance is the \emph{discontinuity} of the accept-reject step, which significantly complicates the argument of sampler verification in the presence of pseudorandom number generation.

\begin{lstlisting}[caption=Rejection sampling, label={alg:rejectionsampling}]
let phi = $\lambda$x : R . 1/sqrt(2*pi) * exp(-1/2*(3-x)*(3-x))
	in let accept = $\lambda$(u,v) : T . 
		if v $\leq$ phi(u)*sqrt(2*pi) then 1 else 0
		in let proj = $\lambda$z : T. fst(cast$\langle$R$\times$R$\rangle$(z)) in
	 		map(proj, reweight(accept, tri $\otimes$ rand))
\end{lstlisting}

In order to show the validity of rejection sampling from the prior, we must assume access to a sampler on the prior distribution (here \(\Code{tri}\)), an independent standard uniform random sampler (here \(\Code{rand}\)), and an upper bound \(\sup_{x \in \R} L(x) = \frac{1}{2\pi} \sup_{x \in \R} \exp(-\frac{(3-x)^2}{2}) = \frac{1}{2\pi}\) on the likelihood, which is used in the acceptance condition.
\cref{fig:rejectionsamplingproof} proves that, subject to the natural independence assumption for the samplers \(\Code{tri}\) and \(\Code{rand}\), the rejection sampler defined by \cref{alg:rejectionsampling} targets the measure \(\denote{\Code{proj}}_* (\denote{\Code{accept}} \cdot (\text{Tri} \otimes U))\). We can then show, using standard methods, that this measure is identical to \(P\), the posterior distribution also targeted by \cref{alg:importancesampling}.

We have omitted, for the moment, one crucial part of the proof. Note that, in both \cref{alg:rejectionsampling} and \cref{fig:rejectionsamplingproof}, the function \(\Code{accept}\), which one might expect to have type \(\Code{R \times R \to R^+}\), instead has type \(\Code{T \to R^+}\). Correspondingly, the product \(\Code{tri \otimes rand}\) must be assumed to produce samples of type \(\Code{T}\), rather than \(\Code{R \times R}\), and \(\Code{proj}\) must accepts inputs of type \(\Code{T}\) rather than pairs \(\Code{R \times R}\). The nature of this type \(\Code{T}\), a \emph{subtype} of \(\Code{R \times R}\), will be explained in \cref{sec:language}, but it encodes the fact that \(\Code{accept}\) is discontinuous when viewed as a function on the standard topologies, as well as where those discontinuities are allowed to lie. The type-inference of \cref{alg:rejectionsampling}, detailing the structure of \(\Code{T}\), is given in \ifarxiv the Appendix, \cref{fig:typecheckingrejection1,fig:typecheckingrejection2}\else\cite[Fig. 6 and Fig. 7]{ourarxivversion}\fi.

\newcommand{\CG}{\mathbf{CG}}
\newcommand{\Top}{\mathbf{Top}}
\newcommand{\cat}{\mathscr{C}}
\newcommand{\Id}{\mathrm{Id}}
\newcommand{\id}{\mathrm{id}}
\newcommand{\beh}{\mathrm{beh}}
\newcommand{\interior}[1]{\mathrm{int}(#1)}
\newcommand{\unfold}{\mathrm{unfold}}
\newcommand{\ev}{\mathrm{ev}}

\section{Language}\label{sec:language}

\subsection{Syntax}\label{sec:syntax}

We use a $\lambda$-calculus with a notion of subtype and
a type constructor \(\stype\) for samplers.  

\subsubsection{Types}

Types are generated by the mostly standard grammar in~\cref{lang:typegrammar},  where the set $\mathrm{Ground}$ of ground types is
\[
\{\Code{N}, \Code{R}, \Code{R}^+\}\cup \{f\inv(i)\mid f\in\{\leq,<,\geq,>,=,\neq\}, i=0,1\}.
\]
Our ground types include the natural, real and nonnegative real numbers, as well as important sets of pairs of reals: for example, $\predinv{<}{1}$ will be denoted,  as the notation suggests,  by the pairs of reals whose first component is strictly smaller than the second.  The boolean type $\Code{B}\defeq 1+1$ will be treated as a ground type.  

The only unusual type constructors are the \emph{pullback types} $\pb{s}{t}{T}$ which -- as the name suggests -- will be interpreted as pullbacks (in fact inverse images), and the \emph{sampler types} $\stype\Code{T}$ which will be defined as the coinductive (stream) types defined by the (syntactic) functors $ \Code{T}\times \Code{R}^+\times-$. In other words, we assume that samplers can be weighted; this covers the special case of unweighted samplers, in which every weight is set to 1. As these are the only coinductive types we need, and to highlight the central role played by samplers, we choose not to add generic coinductive types to the language.

The \emph{subtyping relation} $\sub$ on types is the reflexive transitive closure of the relation generated by the rules of  \cref{lang:subtyping}.

\begin{figure*}[t]
\centering
{
\footnotesize
	\begin{subfigure}{.9\textwidth}
		\begin{center}	
			\begin{align*}
				\Code{S},\Code{T} ::= \Code{G}\in\mathrm{Ground}\mid 1 \mid \Code{S} \times \Code{T} \mid\Code{S} + \Code{T} \mid \pb{s}{t}{T} \mid \Code{S} \to \Code{T} \mid \stype\Code{T}\quad\quad s,t:\Code{T}
			\end{align*}
		\end{center}
		\vspace{-3mm}
		\caption{Type grammar}
		\vspace{3mm}
		\label{lang:typegrammar}
	\end{subfigure}
	
	\begin{subfigure}{.9\textwidth}
		\begin{center}
			\begin{tabular}{l l l l l}
			\multicolumn{3}{l}{
			\infer[f\in\{<,\leq,>,\geq,=,\neq \}]{f\inv(0)+f\inv(1)\sub \Code{R\times R}}{}
			}
			& 
			\infer{\Code{S}_1\times \Code{T}_1\sub\Code{S}_2\times\Code{T}_2}
			{\Code{S}_1\sub\Code{S}_2\qquad\Code{T}_1\sub\Code{T}_2}
			&
			\infer{\Code{S}_1+ \Code{T}_1\sub\Code{S}+\Code{T}_2}
			{\Code{S}_1\sub\Code{S}_2\qquad\Code{T}_1\sub\Code{T}_2}
			\\ \\
			\infer{\stype\Code{S}\sub\stype\Code{T}}{\Code{S}\sub\Code{T}}
			&
			\multicolumn{2}{l}{
			\infer{\sum_{i\in n,j\in m}\Code{S}_i\cap\Code{S}_j'\sub\sum_{i\in n}\Code{S}_i}
			{\sum_{i\in n}\Code{S}_i\sub\Code{S} & \sum_{j\in m}\Code{S}_i'\sub\Code{S}}
			}
			&
			\multicolumn{2}{l}{
			\infer{\sum_{i\in n,j\in m}\Code{S}_i\cap\Code{S}_j'\sub\sum_{j\in m}\Code{S}_j}
			{\sum_{i\in n}\Code{S}_i\sub\Code{S} & \sum_{j\in m}\Code{S}_j'\sub\Code{S}}
			}
			\end{tabular}
		\end{center}
		\vspace{-2mm}
		\caption{Subtyping rules}
		\label{lang:subtyping}
	\end{subfigure}
	\begin{subfigure}{.8\textwidth}
		\begin{center}
			\begin{align*}
				t ::=~ & x\in \text{Var}\mid b\in\{\Code{True}, \Code{False}\}\mid n\in \N\mid r\in \R \mid \qquad & \textit{Variables and constants} \\
				& f(t,\ldots,t), f\in \text{Func}\mid \cast{T}{t}\mid & \textit{Built-in functions} \\
				& \case{t}{x}{s}{n}{i} \mid \ini{i}{t} \mid  \lambda x\Code{:T.}t \mid t(t) \mid \letx{x}{t}{t} \mid & \textit{Programming constructs} \\
				& (t, t) \mid \Code{fst}(t) \mid \Code{snd}(t) \mid & \textit{Products} \\
				& \Code{prng}(t, t) \mid t \otimes t \mid \Code{map}(t, t) \mid \Code{reweight}(t, t) \mid \hd(t)\mid\wt(t) \mid \tl(t) \mid \Code{thin}(t, t) & \textit{Sampler operations}
			\end{align*}
		\end{center}
		\vspace{-2mm}
		\caption{Term grammar}
		\vspace{1mm}
		\label{lang:termgrammar}
	\end{subfigure}
	\caption{Grammars and subtyping rules}
	\vspace{-4mm}
}
\end{figure*}

\subsubsection{Terms} \cref{lang:termgrammar} presents the grammar generating the set \(\text{Expr}\) of terms in our language.  
We assume the existence of a set $\mathrm{Func}$ of built-in functions which come equipped with typing information $f: \Code{T}\to\Code{G}$, where $\Code{G}$ is a ground type. Some built-in functions will be continuous \wrt to the usual topologies, such as the addition operation $+:\Code{R}\times\Code{R}\to\Code{R}$, but others will be discontinuous, such as the comparison operators $\{\leq,<,\geq,>,=,\neq\}:\Code{R}\times\Code{R}\to\Code{B}$.  Dealing with such functions is the main reason for adding coproducts to the grammar, as we will discuss in \cref{sec:denotationalsemantics}.  We also employ the syntactic sugar
\[
\ite{b}{s_{\Code{True}}}{s_{\Code{False}}}\defeq\Code{case}~(b, \_)~\Code{of}~\left\{(i,\_)\Rightarrow s_i\right\}_{i\in\B}.
\]

Most of our language constructs are standard for a typed functional language without recursion, but we endow our language with several nonstandard (sampler) operations:
\begin{itemize}
\item The operation \(\Code{prng}(f, t)\) is used to construct a sampler as a pseudo-random number generator, using an initial value \(t\) and a deterministic endomap \(f\).
\item \(s \otimes t\) represents the product of samplers $s,t$.
\item The syntax \(\Code{map}(f, t)\) maps the function \(f\) over the elements produced by the sampler \(t\) to produce a new sampler, in analogy to the pushforward of a measure.
\item The operation \(\Code{reweight}(f, t)\) applies the reweighting scheme \(f\) to the sampler \(t\) to form a new sampler.
\item Given a sampler \(t\), the operation \(\hd(t)\) returns the first sample produced by \(t\), $\wt(t)$ the weight of the first sample produced by \(t\), and \(\Code{tl}(t)\) returns the sampler \(t\) but with its first sample-weight pair dropped.
\item The operation \(\Code{thin}(n, t)\), given a natural number \(n\) and a sampler \(t\), returns the sampler which includes only those elements of \(t\) whose index is a multiple of \(n\).
\end{itemize}
The intuition and purposes of most of these language constructs was explained in \cref{sec:examples}, and their precise meaning will be made clear when we introduce their semantics.

\subsubsection{Well-formed terms} Our typing system is mostly standard and presented in \cref{lang:typingrules}.  The only non-standard rules are the \emph{context-restriction rule} on the second line of \cref{lang:typingrules}, and the typing rules for the sampler operations, which should be straightforward given their descriptions above. The purpose of the context-restriction rule is, in a nutshell, to be able to pass the result of a computation of type $\Code{T}$ which is continuous \wrt a topology $\tau$ on the denotation of $\Code{T}$, to a computation using a variable of type $\Code{T}$ but which is continuous \wrt to a finer topology $\tau'\supset\tau$ on the denotation of $\Code{T}$.  After application of this rule, it is no longer possible to $\lambda$-abstract on the individual variables of the context. There are good semantic reasons for this feature, which we discuss in \cref{sec:denotationalsemantics}. For readability and intuition's sake, the rule is written using the syntactic sugar
\begin{equation}\label{eq:def_inv_image}
t\inv(\Code{T}_i) \defeq \pb{\cast{T}{\ini{i}{x}}}{t}{T}\qquad\text{where }x:\Code{T}_i
\end{equation}
For the subtyping rules \cref{lang:subtyping}, we use the syntactic sugar
\[
\Code{S}_i\cap\Code{S}_j' \defeq \pb{\cast{S}{\ini{i}{x_i}}}{\cast{S}{\ini{j}{x_j'}}}{S}\quad\text{where }x_i:\Code{S}_i,x_j':\Code{S}_j'
\]

\begin{table*}[t]
	\centering
	{
	\footnotesize
	\begin{tabular}{c c c c}
		\infer[g\in\denote{\Code{G}}]{\Gamma\vdash g : \Code{G}}{} 
		&
		\infer{\Gamma, x:\Code{T},\Delta\vdash x:\Code{T}}{}
		&
		\infer[\mathrm{Func}\ni f : \Code{T} \to \mathtt{G}]
		{\Gamma \vdash f(t) : \Code{G}}
		{\Gamma \vdash t : \Code{T}}
		&
		\infer[\Code{S}\sub\Code{T},\Gamma\sub\Delta]{\Gamma\vdash \cast{T}{t}:\Code{T}}{\Delta\vdash t:\Code{S}}
		\\ \\
		\multicolumn{4}{c}{
		\infer[\sum_{i\in m}\Code{T}_i \sub\Code{T}, \Gamma=x_1:\Code{S}_1, \ldots,x_n:\Code{S}_n]
		{(x_1,\ldots,x_n):\sum_{i\in m} t\inv(\Code{T}_i)\vdash t: \sum_{i\in m}\Code{T}_i}
		{\Gamma\vdash t:\Code{T},}	
		}	
		\\ \\
		\infer{\Gamma \vdash (s, t) : \Code{S\times T}}
		{\Gamma\vdash s : \Code{S}& \Gamma \vdash t : \Code{T}} 
		&
		\infer{\Gamma \vdash \fst(t) : \Code{S}}{\Gamma \vdash t : \Code{S \times T}}
		&
		\infer{\Gamma \vdash \snd(t) : \Code{T}}{\Gamma \vdash t : \Code{S \times T}} 
		&
		\infer{\Gamma \vdash \letx{x}{s}{t}: \Code{T}}{\Gamma, \Code{x:S}  \vdash t : \Code{T}& \Gamma\vdash s : \Code{S} } 
		\\ \\
		\infer
		{\Gamma \vdash \la{x}{\Code{S}}{t : \Code{S} \to \Code{T}}}
		{\Gamma, x:\Code{S}\vdash t : \Code{T}} 
		&
		\infer{\Gamma \vdash t(s) : \Code{T}}{\Gamma \vdash s : \Code{S} & \Gamma \vdash t : \Code{S} \to \Code{T}} 
		& 
		\infer[j\in n]{\Gamma\vdash \ini{j}{t}:\sum_{i\in n}\Code{T}_i}{\Gamma\vdash t:\Code{T}_j}
		&
		\infer{\Gamma\vdash \case{t}{x}{s}{I}{i}:\Code{T}}{\Gamma\vdash t: \sum_{i\in I}\Code{T}_i  & \Gamma, x_i:\Code{T}_i\vdash s_i:\Code{T}}
		\\ \\
		\infer{\Gamma \vdash \hd(t) : \Code{T}}{\Gamma \vdash t : \stype\Code{T}} 
		& 
		\infer{\Gamma \vdash \wt(t) : \Code{R^+}}{\Gamma \vdash t : \stype\Code{T}} 
		&
		\infer{\Gamma \vdash \Code{tl}(t) : \stype\Code{T}}{\Gamma \vdash t : \stype\Code{T}} 
		&
		\infer{\Gamma \vdash s \otimes t : \stype(\Code{S} \times \Code{T})}{\Gamma \vdash s : \stype\Code{S} & \Gamma \vdash t : \stype\Code{T}} 
		\\ \\
		\infer{\Gamma \vdash \Code{prng}(s, t) : \stype\Code{T}}{\Gamma \vdash s : \Code{T} \to \Code{T} & \Gamma \vdash t : \Code{T}} 
		&
		\infer{\Gamma \vdash \Code{thin}(t, n) : \stype\Code{T}}{\Gamma \vdash t : \stype\Code{T} & \Gamma \vdash n : \Code{N}} 
		&
		\infer{\Gamma \vdash \Code{map}(t, s) : \stype\Code{T}}{\Gamma \vdash s : \stype\Code{S} & \Gamma \vdash t : \Code{S} \to \Code{T}} 
		&
		\infer{\Gamma\vdash \Code{reweight}(s,t):\stype \Code{T}}
		{\Gamma\vdash s:\Code{T}\to\Code{R^+} & \Gamma \vdash t:\Code{\stype T }}
	\end{tabular}
	}
	\caption{Typing rules}
	\vspace{-4mm}
	\label{lang:typingrules}
\end{table*}

Our typed lambda calculus does not feature recursion for two reasons. First, it is not necessary: as any computable probability measure can be obtained as a computable pushforward of the uniform measure on the unit interval \cite{huang_morrisett_spitters_2020,hoyrup2009computability}, any sampler language which features the sampler operation $\Code{map}$ can, given a uniform sampler, target any computable probability measure. In particular, many rejection samplers, which are commonly implemented recursively, can alternatively be implemented using the operation \(\Code{reweight}\), as shown in \cref{alg:rejectionsampling}. Second, the categorical semantics of a typed, probabilistic, higher-order lambda calculus with recursion are a very recent area of investigation \cite{vakar2019domain}; we consider the inclusion of recursive samplers to be further work.


\subsection{Operational semantics}\label{sec:operationalsemantics}
In practice, in order to evaluate a program containing a sampler, one must specify a finite number of samples \(N \in \N\) which are to be produced. Our (big-step) operational semantics correspondingly takes the form of a reduction relation \((t, N) \to v\), where the left side consists of a well-typed closed term \(t \in \text{Expr}\) and a number of samples \(N \in \N\), and the right side is a \emph{value} \(v \in \text{Value}\), \ie a term generated by the grammar
\begin{align}
	v ::=~ & x\in \text{Var}\mid g\in\Code{G} \mid (v, v) \mid \ini{i}{v} \mid \lambda x\Code{:T.}~v\label{lang:values}
\end{align}

The rules of this big-step operational semantics, shown in full in \ifarxiv the Appendix, \cref{table:operationalfull}\else\cite[Table VI]{ourarxivversion}\fi, are the usual rules for the standard language constructs, together with additional rules for our implemented sampler operations; these are given in \cref{table:operational}. For notational simplicity, these operations make use of lists \((a, b, c, d)\), which are in fact interpreted within our language as nested pairs \((a, (b, (c, d)))\). In order to keep the rules readable, we also introduce the shorthand $(t,N)\to ((v_1,w_1),\ldots,(v_N,w_N))$ to denote the $N$ reductions

{
\vspace{-2mm}
\small
\begin{align*}
&(\hd(t),\wt(t))\to (v_1,w_1),\\
& (\hd(\tl(t)),\hspace{3pt}\wt(\tl(t)))\to (v_2,w_2), \ldots,\\ 
&(\hd(\tl^{N-1}(t)),\hspace{3pt}\wt(\tl^{N-1}(t)))\to(v_N,w_N).
\end{align*}
}
Note that the product of two weighted samplers has as its weights the product of its factors' weights.  The product and the operation \(\Code{reweight}\) are the only operations modifying the weights of samplers.

The following proposition shows that the operational semantics is well-formed in that for any \(N \in \N\), samplers can only reduce to weighted lists of length \(N\). 

\begin{proposition}\label{prop:opsemvalues}\ifarxiv[\cref{proof:opsemvalues}]\else\cite[Appendix A]{ourarxivversion}\fi~If \(\vdash s : \stype \Code{S}\) is a closed sampler, then for any \(N \in \N\), if \((s, N) \to v\), then \(v\) has the form \(((v_1, w_1), \ldots, (v_N, w_N))\), where \(v_n\) are values and \(w_n \in \Rnn\) are weights. If \(\Code{S}\) is not a sampler type, then \(v_n : \Code{S}\); more generally, each \(v_n\) might be a weighted list itself.
\end{proposition}

\begin{table*}[t]
{
\centering
\footnotesize
\makebox[\textwidth]{%
\begin{tabular}{c c c}
\multicolumn{3}{c}{
\infer{(\Code{map}(s, t), N) \to ((v_1, w_1), \ldots, (v_{N}, w_{N}))}
{\left((s(\hd(t)),\wt(t)), N\right) \to (v_1,w_1) &  \ldots & ((s(\hd(\tl^{N-1}(t)),\wt(\tl^{N-1}(t))), N) \to (v_{N},w_N)}
} 
\\ \\
\multicolumn{3}{c}{
\infer{(\Code{reweight}(s, t), N) \to ((v_1, w_1 ), \ldots, (v_{N}, w_{N}))}
{((\hd(t), s(\hd(t))\cdot \wt(t)),N)\to(v_1,w_1)&\ldots & ((\hd(\tl^{N-1}(t)), s(\hd(\tl^{N-1}(t)))\cdot \wt(\tl^{N-1}(t))),N)\to(v_N,w_N)}
} 
\\ \\
\multicolumn{3}{c}{
\infer{(s \otimes t, N) \to (((v_1, v_1'), w_1 \cdot w_1'), \ldots, ((v_{N}, v_{N}'), w_{N} \cdot w_{N}'))}{(s, N) \to ((v_1, w_1), \ldots, (v_{N}, w_{N})) & (t, N) \to ((v_1', w_1'), \ldots, (v_{N}', w_{N}'))}
} 
\\ \\
\infer{(\Code{hd}(t), N) \to v_1}{(t, N) \to ((v_1, w_1), \ldots, (v_{N}, w_{N}))}
& 
\infer{(\Code{tl}(t), N-1) \to ((v_2, w_2), \ldots, (v_N, w_N))}{(t, N) \to ((v_1, w_1), \ldots, (v_N, w_N))} 
& 
\infer{(\Code{wt}(t), N) \to w_1}{(t, N) \to ((v_1, w_1), \ldots, (v_N, w_N))}
\\ \\
\multicolumn{3}{c}{
\infer{(\Code{thin}(s, t), N) \to ((v_1, w_1), (v_{i+1}, w_{i+1}), (v_{2i+1}, w_{2i+1}), \ldots, (v_{(N-1)i+1}, w_{(N-1)i+1}))}{(s, N) \to i & (t, Ni) \to ((v_1, w_1), \ldots, (v_{Ni}, w_{Ni}))}
}
\\ \\
\multicolumn{3}{c}{
\infer{(\Code{prng}(s, t), N) \to ((v_1, 1), \ldots, (v_{N}, 1))}
{(t, N) \to v_1 & (s(t), N) \to v_2 & \ldots & (s^{N-1}(t), N) \to v_{N}}
}
\end{tabular}
}}
\caption{Big-step operational semantics of sampler operations}
\label{table:operational}
\vspace{-4mm}
\end{table*}

\subsubsection*{The self-product operation}
Having clarified the meaning of the product and of the $\Code{thin}$ operation, we are now in a position to formally justify the operation which we referred to, in \cref{sec:examples}, as the `self-product' of a sampler. To motivate it, consider a sampler \(t:\stype\Code{T}\) which evaluates as \((\Code{t}, 2N) \to (x_1, \ldots, x_{2N})\), where for notational clarity we have omitted the weights. From the above operational semantics, the lagged sampler \(\Code{thin(2, t\otimes tl(t)):\stype (T \times T)}\) evaluates to

{
\vspace{-2mm}
\footnotesize
\begin{flalign*}
(\Code{thin(2, t \otimes tl(t))}, N) \to ((x_1, x_2), (x_3, x_4), \ldots, (x_{2N-1}, x_{2N})).
\end{flalign*}
}
This is the `self-product' which was denoted \(\Code{t}^2\) in \cref{sec:examples}. 
This notion is important because it is the construction which allows us to generate pairs of independent samples from a given sampler. Note that simply taking $t\otimes t$ will produce pairs of perfectly correlated samples: the operational semantics gives $(t\otimes t,N)\to ((x_1,x_1),\ldots(x_N,x_N))$.
More generally, for any \(K \in \N\), we define the $K$-fold self-product of a sampler as
\begin{align}\label{eq:selfproduct}
t^K \defeq \Code{thin}(K , t \otimes \Code{tl}(t) \otimes\ldots\otimes \Code{tl}^{K-1}(t)).
\end{align}
Sampling from $t^K$ is intended to allow the sampling of $K$-tuples of independent deviates generated by the sampler $K$. Ultimately, it is only to define this self-product operation that the sampler operation $\Code{thin}$ is included at all, it being somewhat of an unnatural construct.

\subsection{Denotational semantics}\label{sec:denotationalsemantics}

\subsubsection{Denotational universe}
We will see in \cref{sec:verification} that continuous maps play a special role in the verification of sampler properties.  We therefore need a denotational domain in which continuity is a meaningful concept. We also need a Cartesian closed model, as we want to interpret the lambda-abstraction operation of our calculus. A standard solution is to consider the category of \emph{compactly generated topological spaces}~\cite{steenrod1967convenient,mccord1969classifying,lewis1978stable} (henceforth \emph{CG-spaces}). A topological space $X$ is compactly generated if it is Hausdorff and has the property that $C\subseteq X$ is closed iff $C\cap K$ is closed in $K$ for every compact $K$ in $X$~\cite[\S 1]{steenrod1967convenient}. We need not worry about the theory of these spaces, but the following facts are essential in what follows.

\begin{proposition}[\cite{steenrod1967convenient,lewis1978stable}]\label{prop:cg}
\begin{enumerate}
\item The category $\CG$ of CG-spaces and continuous functions is Cartesian closed.
\item The category $\CG$ is complete and cocomplete.
\item Every metrizable topological space is CG.
\item Locally closed subsets (\ie intersections of an open and a closed subset) of CG-spaces are compactly generated.
\end{enumerate}
\end{proposition}

It is worth briefly describing the Cartesian closed structure of $\CG$. The product is in general different from the product in $\Top$, the category of topological spaces: if the usual product topology is not already compactly generated, then it needs to be modified to enforce compact generation~\cite[\S 4]{steenrod1967convenient}. However, in most practical instances the usual product topology is already compactly generated -- for example, any countable product of metrizable spaces is metrizable, and thus compactly generated by \cref{prop:cg}. The internal hom $[X,Y]$ between CG-spaces $X,Y$ is given by the set of continuous maps $X\to Y$ together with the topology of uniform convergence on compact sets, also known as the compact-open topology~\cite[\S 5]{steenrod1967convenient}.

\subsubsection{Semantics of types}
With this categorical model in place we define the semantics of types. The semantics of ground types is as expected: $\denote{\Code{N}}=\N$, equipped with the discrete topology, and $\denote{\Code{R}}=\R, \denote{\Code{R}^+}=[0, \infty)$ with the usual topology.  The spaces $f\inv(i), f\in\{\leq,<,\geq,>,=,\neq\}, i\in 2$ are interpreted precisely as the notation suggests,  \eg
\begin{align*}
&\denote{<\inv(0)}=\{(x,y)\mid x,y\in\R \wedge x\geq y\},  \\
&\denote{=\inv(1)}=\{(x,x)\mid x\in \R\}
\end{align*}
together with the subspace topology inherited from $\R\times\R$.  Since all these spaces are metrizable, our ground types are interpreted in $\CG$ by \cref{prop:cg}. 

Products (including the unit type) and function types are interpreted in the obvious way using the Cartesian closed structure of $\CG$.  Coproduct types are interpreted by coproducts in $\CG$, and given two terms $s,t:\Code{T}$ interpreted as $\CG$-morphisms $\denote{s}:A\to\denote{\Code{T}}, \denote{t}:B\to\denote{\Code{T}}$, the pullback type $\pb{s}{t}{T}$ is interpreted as the pullback $A\times_{\denote{\Code{T}}} B$ of $\denote{s}$ along $\denote{t}$. All these spaces live in $\CG$ by \cref{prop:cg}. 

Since sampler types are coinductive types, their semantics will hinge on the existence of terminal coalgebras.

\begin{theorem}[Ad\'{a}mek]\label{thm:adamek}
Let $\cat$ be a category with terminal object 1, and $F:\cat\to\cat$ be a functor. If $\cat$ has and $F$ preserves $\omega^{\mathrm{op}}$-indexed limits, then the limit $\nu F$ of \(\xymatrix@C=4ex {1 & F1\ar[l]_{!} & FF1\ar[l]_{F!} & ...\ar[l]_-{FF!}}\) is the terminal coalgebra of $F$.
\end{theorem}

Since $\CG$ is complete, it has $\omega^{\mathrm{op}}$-indexed limits. Recall that we want to interpret $\stype\Code{T}$ as the coinductive type defined by the `functor' $\Code{T}\times\Code{R}^+\times -$. Formally, given a type $\Code{T}$ we want
\begin{align}
\denote{\stype\Code{T}}\defeq \nu (\denote{\Code{T}}\times\R^+\times \Id).\label{eq:sampler_type}
\end{align}
Since products are limits, and limits commute with limits, it is clear that the functor $\denote{\Code{T}}\times\R^+\times \Id$ preserves limits, and in particular $\omega^{\mathrm{op}}$-indexed ones.  Ad\'{a}mek's theorem thus guarantees the existence of an object satisfying \eqref{eq:sampler_type}. More concretely, since the termimal object 1 is trivially metrizable, and since $\R^+$ is metrizable, each object in the terminal sequence will be metrizable provided $\denote{\Code{T}}$ is, and thus $\prod_{n} (\denote{\Code{T}}\times \R^+)^n$ will be metrizable whenever $\denote{\Code{T}}$ is, and will therefore be equipped with the usual product topology. The limit defining \eqref{eq:sampler_type} is a closed subspace of this product, which means that the limit in $\CG$ defining $\denote{\stype\Code{T}}$ is the same as in $\Top$ when $\denote{\Code{T}}$ is metrizable (for example, if $\Code{T}$ is a ground type or a product of ground types).  However, by defining $\denote{\stype\Code{T}}$ coinductively rather than simply as $(\denote{\Code{T}}\times\R^+)^{\omega}$, we obtain a terminal coalgebra structure on $\denote{\stype\Code{T}}$, and therefore the ability to define sampler operations coinductively. 

\subsubsection{Semantics of the subtyping relation}\label{sec:sem_subtype}
Our language contains the predicates \(f\in \{\leq, <, \geq, >, =, \neq\}\) (essential for rejection sampling~\cref{sec:examplerejectionsampling}) and yet is meant to be interpreted in a universe of topological spaces and continuous maps.  These predicates are not continuous maps $\R\times\R\to 2$ for the usual topology on $\R\times\R$. However, for each such predicate $f$, the sets $\denote{f\inv(0)}$ and $\denote{f\inv(1)}$ are \emph{locally closed sets}, that is to say the intersection of an open set and a closed set (for the usual topology on $\R\times\R$), and therefore CG-spaces by~\cref{prop:cg}, \eg $\denote{\predinv{<}{0}}$ is closed and $\denote{\predinv{<}{1}}$ open.

Our central idea for dealing with discontinuities is that since $\CG$ is cocomplete, the space $\denote{f\inv(0)}+\denote{f\inv(1)}$ is a CG-space. This space has the nice property that $f$ \emph{is continuous} as a map $f: \denote{f\inv(0)+f\inv(1)}\to 2$.  Since each $f\inv(i)$ is a type, we can enforce this semantics by simply \emph{typing} these built-in functions in $\mathrm{Func}$ as $f: f\inv(0)+f\inv(1)\to \Code{B}$.

The topology on $\denote{f\inv(0)+f\inv(1)}$ is finer than the usual topology on $\R\times\R$, which means that the identity map $\Id: \denote{f\inv(0)+f\inv(1)}\to \R\times\R$ is continuous. This is the semantic basis for the axiom in~\cref{lang:subtyping}.  
From the other rules it is easy to see by induction that the subtyping relation is always between spaces \emph{sharing the same carrier set} and is semantically given by coarsening the topology. In other words, if $\Code{S}\sub\Code{T}$, then $\denote{\Code{S}}$ and $\denote{\Code{T}}$ share the same carrier and the corresponding identity map $\Id : \denote{\Code{S}}\to\denote{\Code{T}}$ is continuous.

\newcommand{\Eq}{\Code{Eq}}
\newcommand{\Neq}{\Code{Neq}}
\begin{example}\label{ex:single_discont}
Let $\Code{p}\defeq \ite{x=0}{1}{-1}$; we will first show how the context-restriction rule allows us to type-check this program. For readability's sake,  let $\Code{Eq}\defeq \predinv{=}{1}$ and $\Code{Neq}\defeq \predinv{=}{0}$. We now derive, using $=\hspace{1mm}: \Code{Neq+Eq}\to\Code{R}$,

{
\footnotesize
\centering
\[
\infer{x: (x,0)\inv\Neq+(x,0)\inv\Eq\vdash \ite{x=0}{1}{-1}:\Code{R}}
	{\infer{x: (x,0)\inv\Neq +(x,0)\inv\Eq\vdash x=0:\Code{B} \qquad \vdash 1:\Code{R} \qquad \vdash -1:\Code{R}}
		{\infer[\Neq+\Eq\sub\Code{R}\times\Code{R}]
			{\hspace{-12mm}x: (x,0)\inv\Neq +(x,0)\inv\Eq \vdash (x,0):\Neq+\Eq}
				{\infer{x:\Code{R}\vdash (x,0):\Code{R}\times\Code{R}}
				{x:\Code{R}\vdash x:\Code{R}\qquad\vdash 0:\Code{R}}
				}
		}
	}
\]
}
Anticipating the semantics on terms discussed shortly, it can easily be shown that
\begin{gather*}
\denote{(x,0)\inv\Neq +(x,0)\inv\Eq}=
\left((-\infty,0)\cup(0,\infty)\right)+\{0\}
\end{gather*}
and thus $\denote{\Code{p}}$ is the \emph{continuous} map
\[
\denote{\Code{p}}:\left((-\infty,0)\cup(0,\infty)\right)+\{0\}\to\R, x\mapsto 
\begin{cases}
	1 & \text{if }x=0\\
	-1 & \text{else}
\end{cases}
\]
\end{example}

\subsubsection{Semantics of well-formed terms}\label{sec:denotationalsemantics_terms}
Axioms, weakening, subtyping, product, projections, \texttt{let}-binding, $\lambda$-abstraction, function application, injections and pattern matching are interpreted in the expected way (given that $\CG$ is a Cartesian closed category with coproducts).
 
Continuous built-in functions, for example $+:\Code{R}\times\Code{R}\to\Code{R}$ or $\Code{exp}:\Code{R}\to\Code{R}$, are interpreted in the obvious way.  As explained above, discontinuous built-in functions $\{\leq,<,\geq,>,=,\neq\}$ are typed in such a way that their natural interpretations are tautologically continuous. 

We can now describe the semantics of the context-restriction rule.  From the premise,  our observations in \cref{sec:sem_subtype}, and the side-conditions, we have morphisms 
\[
\denote{t}:\prod_{j \in n}\denote{\Code{S}_j}\to\denote{\Code{T}},\quad\text{and}\quad\Id: \coprod_{i\in m}\denote{\Code{T}_i}\to\denote{\Code{T}}.
\]
By \cref{eq:def_inv_image} we interpret each `inverse image type' $t\inv(\Code{T}_i)$ as the pullback (inverse image) of $\denote{t}$ along the inclusion $\denote{\Code{T}}_i\hookrightarrow \coprod_{i\in m}\denote{\Code{T}_i}$ which is, as the notation implies, simply given by $\denote{t}\inv(\denote{\Code{T}_i})$. Since $\coprod_{i\in m}\denote{\Code{T}_i}$ and $\denote{\Code{T}}$ share the same carrier, it is clear that this defines a partition of $\denote{\Gamma}$, and we can thus retype $t$ as a continuous map $\coprod_{i\in m}\denote{t\inv(\Code{T}_i)} \to \coprod_{i\in m}\denote{\Code{T}_i}$, interpreting the rule.

As mentioned earlier in this section, context-restriction prevents $\lambda$-abstraction; the following example illustrates why this must be the case.

\begin{example}
Consider the program $x<y$ derived by:
\[
{\footnotesize
\infer
{(x,y):(x,y)\inv(\predinv{<}{0})+(x,y)\inv(\predinv{<}{1})\vdash x<y:\Code{B}}
	{\infer{(x,y)\hspace{-2pt}:\hspace{-2pt}(x,y)\inv(\predinv{<}{0})+(x,y)\inv(\predinv{<}{1})\vdash (x,y)\hspace{-2pt}:\hspace{-2pt}\predinv{<}{0}+\predinv{<}{1}}
		{x:\Code{R},y:\Code{R}\vdash (x,y):\Code{R}\times\Code{R}}
	}
}
\]
The interpretation of $x<y$ is given by the continuous function 
\[
\denote{<}:\{(x,y)\mid x<y\}+\{(x,y)\mid x\geq y\}\to 2.
\]
Although it has the same carrier $\R\times\R$, the domain of this map is no longer of product of topological spaces; it is now a coproduct of topological spaces. This means that it is no longer possible to $\lambda$-abstract over one of the variables of this function using the Cartesian closed structure of $\CG$.

In order to be able to $\lambda$-abstract the map $<$, we would need a topology on $\R\times\R$ with the property that for any given $x_0\in \R$ the function $x_0<-: \R\to 2$ is continuous.  This would introduce the open sets $[x_0,\infty)$ to the topology of $\R$ for each $x_0\in\R$, meaning that we must equip $\R$ with the notoriously problematic lower limit topology (\aka the Sorgenfrey line).  Whether or not this is a CG-space seems to be a thorny question, possibly independent of ZF \cite{keremedis2018compact}.
\end{example}

Finally, we define the denotational semantics of sampler operations using the coinductive nature of sampler types. Recall that for a type $\Code{T}$,  $\denote{\stype\Code{T}}\defeq \nu (\denote{\Code{T}}\times\R^+\times \Id)$.  In particular, $\denote{\stype\Code{T}}$ comes equipped with a coalgebra structure map
\[
\mathrm{unfold}_{\Code{T}}: \denote{\stype\Code{T}}\to \denote{\Code{T}}\times \R^+\times \denote{\stype\Code{T}}.
\]
Moreover, for any other (continuous) coalgebra structure map $\gamma: X\to \denote{\Code{T}}\times \R^+\times X$, the terminal nature of $\denote{\stype\Code{T}}$ provides a unique $\denote{\Code{T}}\times \R^+\times\Id$-coalgebra morphism
\[
\beh(\gamma): X\to \denote{\stype\Code{T}}.
\]
Since $\denote{\stype\Code{T}}$ is interpreted in $\CG$, it follows automatically that both $\mathrm{unfold}_{\Code{T}}$ and $\beh(\gamma)$ are continuous.  However, what is not immediately clear is that $\beh$ is in fact continuous in $\gamma$.

\begin{proposition}\ifarxiv[\cref{proof:continuousbeh}]\else\cite[Appendix A]{ourarxivversion}\fi\label{prop:beh}~Let $F: \CG\to\CG$ satisfy the condition of \cref{thm:adamek} as well as the condition that $\interior{\nu F}\neq\emptyset$ in $\prod_i F^i 1$,  and let $\beh_X: [X,FX]\to [X, \nu F]$ be the (behaviour) map associating to any $F$-coalgebra structure on $X$ the unique coalgebra morphism into the terminal coalgebra. The map $\beh_X$ is continuous, \ie is a $\CG$-morphism.
\end{proposition}
Using $\unfold$ and $\beh$ we define the denotational semantics of all the sampler operations in \cref{table:denotationalsemantics}. These definitions are precisely the infinite (coinductive) versions of the finitary transformations defined in the operational semantics of \cref{table:operational}. All the maps involved in these definitions are continuous; this follows from \cref{prop:beh} and the fact that evaluation and function composition are continuous operations on the internal hom sets of $\CG$ (\cite[5.2,5.9]{steenrod1967convenient}).

\begin{table*}[h]
\centering
\begin{tabular}{c c c c}
\infer{\denote{\Gamma \vdash \Code{hd}(t) : \Code{T}} = \pi_1\circ \mathrm{unfold}_{\Code{T}}\circ f}{\denote{\Gamma \vdash t : \stype\Code{T}} = f} 
& 
\infer{\denote{\Gamma \vdash \Code{wt}(t) : \Code{R}^+} = \pi_2 \circ \mathrm{unfold}_{\Code{T}} \circ f}{\denote{\Gamma \vdash t : \stype \Code{T}} = f}
& 
\multicolumn{2}{c}{
\infer{\denote{\Gamma \vdash \Code{tl}(t) : \stype\Code{T}} = \pi_3\circ \mathrm{unfold}_{\Code{T}}\circ f}{\denote{\Gamma \vdash t : \stype\Code{T}} = f} 
}
\\ \\
\multicolumn{4}{c}{
\infer{\denote{\Gamma \vdash \Code{thin}(s, t) : \stype(\Code{T})} =\ev_{\stype\Code{T},\stype\Code{T}}\circ
(\id_{\stype\Code{T}}\times \beh_{\stype\Code{T}})\circ 
\left(\id_{\stype\Code{T}}\times \left(\unfold_{\Code{T}} \circ (\pi_3\circ\unfold_{\Code{T}})^{(\cdot~-1)}\right)\right)\circ \langle f,g\rangle
}
{\denote{\Gamma \vdash s : \Code{N}} = f & \denote{\Gamma \vdash t : \stype\Code{T}} = g}
} 
\\ \\
\multicolumn{4}{c}{
\infer{\denote{\Gamma \vdash s \otimes t : \stype(\Code{S} \times \Code{T})} = \beh_{\stype\Code{S},\stype\Code{T}}\left(\pi_1\times\pi_4\times (\pi_2\cdot\pi_5)\times\pi_3\times\pi_6\circ\mathrm{unfold}_{\Code{S}}\times \mathrm{unfold}_{\Code{T}}\right)\circ \langle f, g\rangle}
{\denote{\Gamma \vdash s : \stype\Code{S}} = f & \denote{\Gamma \vdash t : \stype\Code{T}} = g}
}
\\ \\
\multicolumn{4}{c}{
\infer{\denote{\Gamma \vdash \Code{map}(t, s) : \stype\Code{T}} = \ev_{\stype\Code{S},\stype\Code{T}}\circ
(\id_{\stype\Code{S}}\times \beh_{\stype\Code{S}})\circ 
\left(\id_{\stype\Code{S}}\times ((-\times\id_{\Code{R}^+}\times\id_{\stype\Code{S}})\circ\unfold_{\Code{S}})\right) \circ \langle f,g\rangle}
{\denote{\Gamma \vdash s : \stype \Code{S}} = f & \denote{\Gamma \vdash t : \Code{S} \to \Code{T}} = g}
} 
\\ \\
\multicolumn{4}{c}{
\infer{\denote{\Gamma \vdash \Code{reweight}(t, s)} = \ev_{\stype\Code{T},\stype\Code{T}}\circ
(\id_{\stype\Code{T}}\times \beh_{\stype\Code{T}})\circ 
\left(\id_{\stype\Code{T}}\times ((\id_{\Code{T}}\times -\times\id_{\stype\Code{T}})\circ\unfold_{\Code{T}})\right) \circ \langle f,g\rangle}
{\denote{\Gamma \vdash s : \stype\Code{T}} = f & \denote{\Gamma \vdash t : \Code{T} \to \Code{R}^+} = g}
}
\\ \\
\multicolumn{4}{c}{
\infer{\denote{\Gamma \vdash \Code{prng}(s, t) : \stype\Code{T}} = 
\ev_{\Code{T},\stype\Code{T}}\circ (\id_{\Code{T}}\times \beh_{\Code{T}})\circ \left(\id_{\Code{T}}\times(\id_{\Code{T}}\times 1\times -)\right)\circ \langle f,g\rangle
}
{\denote{\Gamma \vdash t : \Code{T}} = f &  \denote{\Gamma \vdash s : \Code{T} \to \Code{T}} = g }
} 
\end{tabular}
\caption{Denotational semantics of sampler operations}
\vspace{-4mm}
\label{table:denotationalsemantics}
\end{table*}

\subsection{Adequacy}

This language features an interesting asymmetry in that its denotational semantics is written in terms of the coinductive sampler type \(\denote{\stype\Code{T}}\), while its operational semantics is written in terms of finitary operations on finite sequences of samples.  Moreover, the operational semantics is given in terms of reductions to \emph{values}, \ie terms whose types are constructed without the type constructor $\stype$, whereas the denotational semantics does not make this distinction. To establish a connection, we start by defining a generic way to convert terms of arbitrary types into values, following the idea behind the operational semantics.
Given a type $\Code{T}$ and an integer $N$ we inductively define its associated value type $\mathrm{val}^N(\Code{T})\in\mathrm{Value}$ by:
\begin{flalign*}
{
\small
\begin{tabular}{l r}
$\mathrm{val}^N({\Code{G}})=\Code{G}$
&
$\qquad \mathrm{val}^N(\stype\Code{T})=\left(\mathrm{val}^N\Code{T}\right)^N$
\\ 
\multicolumn{2}{l}{
$\mathrm{val}^N({\Code{S} \ast \Code{T}})=\mathrm{val}^N(\Code{S}) \ast \mathrm{val}^N(\Code{T}), \quad \ast\in\{\times,+,\to\}$
}
\end{tabular}
}
\end{flalign*}
where $\Code{G}\in\mathrm{Ground}$.\footnote{Since we're only interested in closed samplers here, and since pullback types can only occur in a context, we need not define $\mathrm{val}^N$ on pullback types.} We now define the \emph{generalized projection} maps $\pnt: \denote{\Code{T}}\to \denote{\mathrm{val}^N(\Code{T})}$ recursively via
\begin{flalign*}
{
\small
\begin{tabular}{l l}
$\pnt[G]=\id_{\denote{\Code{G}}},$
&
$\pnt[S\ast T]=\pnt[S]\ast \pnt, \ast\in\{\times,+\}$
\\
$\pnt[S\to T]=\id_{\denote{\Code{S\to T}}}$
&
$\pnt[\stype T]=\pi_{1:N}\circ (\pnt[T\times R^+])^\omega$
\end{tabular}
}
\end{flalign*}
The reader will have noticed that we have defined $\pnt[S\to T]$ trivially. The reason is that,  as a quick examination of the rules of \cref{table:operational} will reveal,  there is no conclusion and no premise of the type $(t,N)\to v$ where $t$ is of function type. The only occurrence of terms of function types are within an evaluation, or are \emph{values}, \ie terms trivially reducing to themselves.

\begin{theorem}\ifarxiv[\cref{proof:adequacy}]\else\cite[Appendix A]{ourarxivversion}\fi\label{thm:adequacy}~For any program  $\vdash t:  \Code{T}$, we have
\[
(t,N) \to v \Leftrightarrow\pnt(\denote{t}) = \denote{v}.
\]
\end{theorem}


\section{Equivalence of samplers}\label{sec:equivalence}

In order to implement a system for reasoning about whether a deterministic sampler targets a particular probability distribution, it is necessary to first define a notion of equivalence between samplers. Having such a system gives a natural path towards verifying a sampler: first rewrite a given sampler \(s\) in an equivalent but simpler form, and then show that this simplified form targets the correct distribution. This is the approach taken in the derivations in \cref{sec:examples}, which implicitly used several equivalence results -- in particular, \(\Code{let}\)-reduction and the equivalence of the nested self-product \((s^m)^n\) to the self-product \(s^{m*n}\) for any sampler \(s\). In this section, we introduce a relation \(\approx\) on programs which justifies this type of reasoning.

\begin{definition}
We say that two programs \(\Gamma \vdash s:\Code{T}\) and \( \Gamma \vdash t: \Code{T}\) are \emph{equivalent}, notation \(\Gamma\vdash s\approx t: \Code{T}\),  if they are related by the smallest congruence relation on well-typed terms containing the rules of \cref{table:equivalencerules}.\footnote{By \emph{congruence relation}, we mean that $\approx$ is an equivalence relation preserved by all operations in the language. For example, if \(\Gamma\vdash s\approx t:\stype \Code{T}\) holds, then \(\Gamma\vdash \Code{tl}(s)\approx \Code{tl}(t):\stype \Code{T}\) must hold as well, and the same for all operations in the language.}
\end{definition}

The rules of \cref{table:equivalencerules} employ a number of shorthand conventions for a more concise presentation. We introduce identity functions \(\Code{id_S} \defeq \lambda x:\Code{S}.~ x :\Code{S \to S}\), constant functions \(\Code{1_S} \defeq \lambda x:\Code{S. ~1 : S \to R^+}\),  function composition \(t \circ s \defeq \lambda x:\Code{S}.~ t(s(x)) : \Code{S \to U}\) where  \(s : \Code{S \to T}, t:\Code{T \to U}\), compositions \(\Code{f^0 \defeq id_S : S \to S, f^n \defeq f \circ f^{n-1}}\) for any \(n \in \N\), pointwise products \(s \cdot t \defeq \lambda x : \Code{S}, y : \Code{T}.~s(x) * t(y) : \Code{S \times T \to R^+}\) of real-valued functions \(s : \Code{S \to R^+}, t : \Code{T \to R^+}\), and finally Cartesian products \(s \times t \defeq \lambda x : \Code{S}, y : \Code{T}. ~(s(x), t(y)) : \Code{S \times T \to S' \times T'}\) of functions \(s : \Code{S \to S'}, t :\Code{T \to T'}\).

\begin{table*}[t]
\centering

\small{
\begin{tabular}{c c}

$\Gamma, s:\Code{S} \vdash (\lambda x:\Code{S} . t)(s) \approx t[x \leftarrow s] : \Code{T}$ & $\Gamma \vdash \lambda x:\Code{S}.t(x) \approx t : \Code{S \to T}$ \\[3pt]
\multicolumn{2}{c}{$\Gamma, s:\Code{S} \vdash \letx{x}{s}{t}\approx (\lambda x:\Code{S}.~t)(s):\Code{T}$} \\[3pt]
$\Gamma \vdash \ite{\Code{True}}{s}{t} \approx s:\Code{T}$ & $\Gamma \vdash\ite{\Code{False}}{s}{t}\approx t : \Code{T}$ \\[3pt]
$\Gamma \vdash \Code{fst}((s, t)) \approx s:\Code{S}$ & $\Gamma \vdash \Code{snd}((s, t)) \approx t : \Code{T}$
\end{tabular}
}


\vspace{2ex}

\footnotesize{
\begin{tabular}{c c c}

$\Gamma \vdash \Code{hd(map}(s, t)) \approx s(\Code{hd}(t)) : \Code{T}$ & $\Gamma \vdash \Code{wt}(\Code{map}(s, t)) \approx \Code{wt}(t) : \Code{R^+}$ & $\Gamma \vdash \Code{tl(map}(s, t)) \approx \Code{map}(s, \Code{tl}(t)) : \stype \Code{T}$ \\[3pt]
$\Gamma \vdash (\Code{hd}(s), \Code{hd}(t)) \approx \Code{hd}(s \otimes t) : \Code{S \times T}$ & $\Gamma \vdash \Code{wt}(s)*\Code{wt}(t) \approx \Code{wt}(s \otimes t) : \Code{R^+}$ & $\Gamma \vdash \Code{tl}(s) \otimes \Code{tl}(t) \approx \Code{tl}(s \otimes t) : \stype \Code{(S \times T)}$ \\[3pt]
$\Gamma \vdash \Code{hd(thin}(n, t)) \approx \Code{hd}(t) : T$ & $\Gamma \vdash \Code{wt}(\Code{thin}(n, t)) \approx \Code{wt}(t) : \Code{R^+}$ & $\left\{\Gamma \vdash \Code{tl(thin}(n, t)) \approx \Code{thin}(n, \Code{tl}^n(t)) : \stype \Code{T} \mid n \in \N\right\}$ \\[3pt]
& $\Gamma \vdash \Code{thin(1}, t) \approx t : \stype \Code{T}$ & \\[3pt]
$\Gamma \vdash \Code{hd(prng}(s, t)) \approx t : \Code{T}$ & $\Gamma \vdash \Code{wt}(\Code{prng}(s, t)) \approx 1 : \Code{R^+}$ & $\Gamma \vdash \Code{tl(prng}(s, t)) \approx \Code{prng}(s, s(t)) : \stype \Code{T}$ \\[3pt]
$\Gamma \vdash \Code{hd(reweight}(s, t)) \approx \Code{hd}(t) : \Code{T}$ & $\Gamma \vdash \Code{wt}(\Code{reweight}(s, t)) \approx s(\Code{hd}(t)) * \Code{wt}(t) : \Code{R^+}$ & $\Gamma \vdash \Code{tl(reweight}(s, t)) \approx \Code{reweight}(s, \Code{tl}(t)) : \stype \Code{T}$

\end{tabular}
}

\vspace{2ex}

\small{
\begin{tabular}{c c}

$\Gamma \vdash \Code{thin}(n, \Code{thin}(m, t)) \approx \Code{thin}(n*m,t) \approx \stype \Code{T}$ & $\Gamma \vdash \Code{map}(g, \Code{map}(f, t)) \approx \Code{map}(g \circ f, t) : \stype \Code{T}$ \\[3pt]
\multicolumn{2}{c}{$\Gamma \vdash \Code{reweight}(g, \Code{reweight}(f, t)) \approx \Code{reweight}(f \cdot g, t) : \stype \Code{T}$} \\[3pt]
$\left\{\Gamma \vdash \Code{thin}(n, \Code{prng}(s, t)) \approx \Code{prng}(s^n, t) : \stype \Code{T} \mid n \in \N\right\}$ & $\Gamma \vdash \Code{thin}(n, \Code{map}(s, t)) \approx \Code{map}(s, \Code{thin}(n, t)) : \stype \Code{T}$ \\[3pt]

$\Gamma \vdash s \otimes \Code{map}(f, t) \approx \Code{map(id_S} \times f, s \otimes t) : \stype \Code{(S \times T)}$
&
$\Gamma \vdash \Code{map}(f, s) \otimes t \approx \Code{map}(f \times \Code{id_T}, s \otimes t) : \stype \Code{(S \times T)}$ \\[3pt]
$\Gamma \vdash s \otimes \Code{reweight}(g, t) \approx \Code{reweight(1_S} \cdot g, s \otimes t) : \stype \Code{(S \times T)}$ &
$\Gamma \vdash \Code{reweight}(f, s) \otimes t \approx \Code{reweight}(f \cdot \Code{1_T}, s \otimes t) : \stype \Code{(S \times T)}$ \\[3pt]
$\Gamma \vdash \Code{prng}(f, a) \otimes \Code{prng}(g, b) \approx \Code{prng}(f \times g, (a, b)) : \stype \Code{(S \times T)}$ & 
$\Gamma \vdash \Code{thin}(n, s) \otimes \Code{thin}(n, t) \approx \Code{thin}(n, s \otimes t) : \stype \Code{(S \times T)}$

\end{tabular}
}

\caption{Rules for sampler equivalence}
\vspace{-4mm}
\label{table:equivalencerules}
\end{table*}

\begin{theorem}\label{thm:sound_equivalence}\ifarxiv[\cref{proof:sound_equivalence}]\else\cite[Appendix A]{ourarxivversion}\fi~The rules of \cref{table:equivalencerules} are sound: if $\Gamma\vdash s\approx t:\Code{T}$, then $\denote{\Gamma \vdash s:\Code{T}}=\denote{\Gamma \vdash t:\Code{T}}$.
\end{theorem}
\noindent The proof is a straightforward exercise in coinductive reasoning and can be found in the Appendix, along with the full list of equivalence rules.
The soundness of these rules with respect to operational equivalence then follows from abstraction, though it is also straightforward to show directly.

Recall that an important application of our sampler operations is to provide a formal definition of the \emph{self-product} of samplers, given in \eqref{eq:selfproduct}. It is crucial that our equivalence rules should show that this self-product is well-defined.

\begin{proposition}\label{prop:nestedselfproduct}\ifarxiv[\cref{proof:nestedselfproduct}]\else\cite[Appendix A]{ourarxivversion}\fi~For any \(\Gamma \vdash s : \stype \Code{S}\), \(m, n \in \N\), the self-product satisfies \(\Gamma \vdash (s^m)^n \approx s^{mn} : \stype (\Code{S}^{mn})\).
\end{proposition}

The equivalence rules in \cref{table:equivalencerules} suggest a procedure for simplifying samplers. Consider samplers which have no occurrences of the operation \(\Code{prng}\). Of our remaining sampler operations, we identify two groups: \(\{\Code{tl}, \Code{hd}, \Code{wt}, \Code{thin}, \otimes\}\) and \(\{\Code{map}, \Code{reweight}\}\). Applying the rules of \cref{table:equivalencerules}, we see that for each combination of operations in the first and second group, there is a rule which enables us to pull the first operation into the body of the second. Therefore, any sampler with no instances of \(\Code{prng}\) can be written so that the sampler operations \(\Code{tl}, \Code{hd}, \Code{wt}, \Code{thin}, \otimes\) are pulled all the way inwards.

\begin{proposition}
Let \(\Gamma \vdash t : \stype \Code{T}\) be a sampler which contains no instances of \(\Code{prng}\). Applying the rules of \cref{table:equivalencerules}, it follows that we can equivalently rewrite such a sampler in either the form \(\Gamma \vdash \Code{map}(f, \Code{reweight}(g, \Code{map}(f', \ldots, s))) : \stype \Code{T}\) or \(\Gamma \vdash \Code{reweight}(g, \Code{map}(f, \Code{reweight}(g', \ldots, s))) : \stype \Code{T}\), \ie a composition of invocations of \(\Code{map}\) and \(\Code{reweight}\) (including the trivial case of zero occurrences of either), where crucially the sampler \(s\) does not contain the sampler operations \(\Code{map}\) or \(\Code{reweight}\).
\end{proposition}

We can also show that the self-product distributes over the operations \(\Code{map}\) self-and \(\Code{reweight}\); such operations are useful for representing the self-product of a composite sampler in a simpler form whose correctness can then be verified.

\begin{proposition}\label{prop:mapreweightselfproduct}\ifarxiv[\cref{proof:mapreweightselfproduct}]\else\cite[Appendix A]{ourarxivversion}\fi~For any mapped sampler \(\Gamma \vdash \Code{map}(f, s) : \stype \Code{T}\) and any \(n \in \N\), it follows that \(\Gamma \vdash \Code{map}(f, s)^n \approx \Code{map}(f \times \ldots \times f, s^n) : \stype (\Code{T}^n)\); for a reweighted sampler \(\Gamma \vdash \Code{reweight}(f, s) : \stype \Code{S}\), it follows that \(\Gamma \vdash \Code{reweight}(f, s)^n \approx \Code{reweight}(f \cdot \ldots \cdot f, s^n) : \stype (\Code{S}^n)\).
\end{proposition}

\section{Semantic correctness of samplers}\label{sec:verification}

The fundamental correctness criterion for a sampler is that it should produce samples which are distributed according to the desired target distribution. This section aims to sketch a simple `targeting calculus' to compositionally verify this property. We frame this correctness in terms of weak convergence of measures; while other notions of convergence could be used, weak convergence is standard and will suffice for our purposes.

\subsection{The empirical transformation}\label{sec:empiricaltransformation}

First, we need to formalise what we mean when we say that a sampler $s:\stype\Code{T}$ \emph{targets} a probability distribution on $\denote{\Code{T}}$. 
Given a topological space $X$, let us write $\pr X$ for the space of probability measures on (the Borel $\sigma$-algebra generated by) $X$, equipped with the topology of weak convergence, \ie $\lim_{n\to\infty}\mu_n=\mu$ in $\pr X$ if for any \emph{bounded continuous} map $f: X\to \R$, $\lim_{n\to\infty}\int f~d\mu_n=\int f~d\mu$. 
In fact, $\pr$ defines a functor $\Top\to\Top$: if $f: X\to Y$ is a continuous map, then $\pr(f)\defeq f_{\ast}:\pr X\to\pr Y$ is the pushforward map, which is easily shown to be continuous.
We do not know if $\pr X$ is a CG-space when $X$ is, and in particular we do not know if $\pr$ can be given a monad structure on $\CG$. These questions are, however, orthogonal to this work since $\pr$ plays no role in the semantics of \cref{sec:denotationalsemantics}.

For any stream $\sigma: \N\to X\times \R^+$ we define $\hat{\sigma}_n\in\pr X$ as
\[
\hat{\sigma}_n\defeq \frac{1}{n}\sum_{i=1}^n\frac{\pi_2(\sigma(i))}{\sum_{j=1}^n \pi_2(\sigma(j))}\delta_{\pi_1(\sigma(i))},
\]
the \emph{empirical distribution} based on the first $n$ (weighted) samples of $\sigma$.  We also define $\prb X\defeq \pr X+1$, where $1=\{\bot\}$ is the terminal object and $+$ the coproduct in $\Top$.

\begin{definition}
The \emph{empirical measure transformation}  is the $\Top^{\mathrm{obj}}$-collection of maps
\[
\varepsilon_X: (X\times \R^+)^\N\to\prb X,  \sigma\mapsto
\begin{cases}
\lim\limits_{n\to\infty} \hat{\sigma}_n &\text{if it exists}\\
\bot\in 1&\text{else}
\end{cases}
\]
\end{definition}

The empirical measure transformation cannot be natural, as the following example shows.

\begin{example}
Let $\sigma:\N\to X$ be a diverging unweighted sampler on $X$, \ie $\varepsilon_X(\sigma)=\bot$, and consider the map to the terminal object $!: X\to 1$. Then \(\varepsilon_1(! \circ \sigma) = \varepsilon_1(\bot, \bot, \ldots) = \delta_{\bot}\), but \(\pr_{\bot}(!)(\varepsilon_X(\sigma)) = \pr_{\bot}(!)(\bot)= \bot\).
\end{example}

However, if a sampler does define a probability measure via $\varepsilon$, this is preserved by continuous maps.

\begin{proposition}\label{prop:continuous_case}\ifarxiv[\cref{proof:continuous_case}]\else\cite[Appendix A]{ourarxivversion}\fi~Let $\sigma:\N\to X\times \R^+$ and $f: X\times\R^+\to Y\times \R^+$ be continuous. If $\varepsilon_X(\sigma)=\mu$, then $\varepsilon_Y(f\circ \sigma)=f_\ast(\mu)$.
\end{proposition}


It is tempting to try to generalise this nice property of continuous maps to more general maps -- for example, measurable maps. The following example shows that this is not possible.
\begin{example}
Let $X=[0,1]$ and $\sigma:\N\to[0,1]$ denote any unweighted sampler such that $\varepsilon(\sigma)$ is the Lebesgue measure on $[0,1]$. Now consider the map $f : [0,1]\to \{0,1\}$ defined by $f(x)=1$ if $x=\sigma(i)$ for some $i$ and 0 else.
This function is the indicator function of a countable, therefore closed, set, and so is Borel-measurable.
On the one hand we have that $\varepsilon(f\circ \sigma)=\delta_1$ since $f\circ \sigma$ is the constant stream on ones, but on the other we have
$f_\ast(\varepsilon(\sigma))(1) = \varepsilon(\sigma)(f\inv(1))=0$ 
since only a countable set is mapped onto 1 by $f$.
\end{example}
Even for functions with finitely many discontinuities, it is impossible to extend the class of functions for which \cref{prop:continuous_case} holds. However, the semantic framework adopted in \cref{sec:denotationalsemantics} allows us to bypass this problem altogether. We illustrate these two points by revisiting \cref{ex:single_discont}.

\begin{example}\label{ex:discont0}
Consider the sampler $\Code{s}\defeq\Code{prng}(\la{x}{\Code{R}}{x/2},1)$ and the term $\Code{p}\defeq \ite{x=0}{1}{-1}$ of \cref{ex:single_discont}. 
Assume first that $\R$ is equipped with its standard topology, \ie that $\denote{\Code{p}}$ is not continuous at 0.  Since $\R$ is a metric space we can use the Portmanteau Lemma and rephrase weak convergence by limiting ourselves to bounded Lipschitz functions. It is then easy to show that $\varepsilon(\denote{\Code{s}})=\delta_0$: letting $f:\R\to\R$ be bounded Lipschitz, we have
\begin{align*}
\lim_{n\to\infty} \left\lvert\int f~d\widehat{\denote{\Code{s}}}_n - \int f~d\delta_0\right\rvert = \lim_{n\to\infty}\left\lvert\frac{1}{n}\sum_{i=1}^n f\left(\frac{1}{2^i}\right) - f(0)\right\rvert \\
\leq \lim_{n\to\infty}\left\lvert\frac{1}{n}\sum_{i=1}^n \frac{1}{2^i}\right\rvert \leq \lim_{n\to\infty}\frac{2}{n} = 0
\end{align*}
\cref{prop:continuous_case} now fails on $\denote{\Code{p}}$, since $\varepsilon(\denote{\Code{p}}\circ \denote{\Code{s}})=\varepsilon(-1,-1,\ldots)=\delta_{-1} \neq \pr(\denote{\Code{p}})(\varepsilon(\denote{\Code{s}}))=\denote{\Code{p}}_\ast(\delta_0)=\delta_1$.

Let us now equip $\R$ with the topology given by type-checking $\Code{p}$ as described in \cref{ex:single_discont}. This makes $\denote{\Code{p}}$ bounded and continuous, and we therefore no longer have $\varepsilon(\denote{\Code{s}})=\delta_0$; indeed $\lim_n \int \denote{\Code{p}}~d\widehat{\denote{\Code{s}}}_n = -1 \neq \denote{\Code{p}}(0)=1$. In fact we now have $\varepsilon(\denote{\Code{s}})=\bot$, \ie $\Code{s}$ is no longer a sampler targeting anything for this topology, which prevents the failure of \cref{prop:continuous_case} on $\denote{\Code{p}}$.
\end{example}

This example also shows that our semantics has provided us with many more morphisms satisfying \cref{prop:continuous_case} than would have been the case had we only considered programs which are continuous \wrt the usual topology on the denotation of types. Our semantics allows us to push forward a sampler \(s\) through any piecewise continuous function, except in the narrow case where this function has a point of discontinuity which is asymptotically assigned positive mass by \(s\). We illustrate this further in the next example.

\begin{example}\label{ex:discont2}
Consider the sampler of \cref{ex:discont0}, but now let $\Code{p}\defeq \ite{x=2}{1}{-1}$ instead.  To make this function continuous, our semantics adds the open set $\{2\}$ to the usual topology of $\R$.  This does not interfere with the derivation that $\varepsilon(\denote{\Code{s}})=\delta_0$, since we can write \(\int f~d\widehat{\denote{\Code{s}}}_n = \int_{\{2\}^c}f~d\widehat{\denote{\Code{s}}}_n+\int_{\{2\}}f~d\widehat{\denote{\Code{s}}}_n = \int_{\{2\}^c}f~d\widehat{\denote{\Code{s}}}_n\), and similarly for $\delta_0$.  Because the discontinuity of $\denote{\Code{p}}$ is not assigned any mass by $\delta_0$, the topology on $\R$ making $\denote{\Code{p}}$ continuous no longer prevents $\varepsilon(\denote{\Code{s}})$ from converging, and we can therefore safely push $\Code{s}$ forward through $\Code{p}$ using $\map$.
\end{example}

\subsection{Calculus for asymptotic targeting}\label{sec:asymptotictargeting}


\begin{table*}[t]
\center
{
\footnotesize
\begin{tabular}{c c c c c c}
\multicolumn{2}{c}{
\infer[i \in I]{\Gamma_i \vdash \Code{rand}_i : \stype \Code{T}_i\rightsquigarrow \mu_i}{}
}
& 
\multicolumn{2}{c}{
\qquad
\infer{\Gamma\vdash t:\stype\Code{ T}\rightsquigarrow \mu}
{\Gamma\vdash s \approx t:\Code{\stype T} & \Gamma\vdash s:\stype \Code{T}\rightsquigarrow \mu} 
}
& 
\multicolumn{2}{c}{
\qquad
\infer{\Gamma\vdash\Code{tl(}s): \stype \Code{S}\rightsquigarrow \mu}
{\Gamma\vdash s:\stype \Code{S}\rightsquigarrow \mu} 
}
\\ \\
\multicolumn{3}{c}{
\infer{\Gamma\vdash\Code{map}(f,s):\stype \Code{T}\rightsquigarrow \gamma \mapsto (\denote{f}(\gamma))_\ast\mu(\gamma)}
{\Gamma\vdash s:\stype \Code{S} \rightsquigarrow \mu & \Gamma \vdash f:\Code{S \to T}}
} 
&
\multicolumn{3}{c}{
\qquad
\infer[\int_{\denote{S}} \denote{f}(\gamma)\,d\mu(\gamma) \in (0, \infty)]
{\Gamma\vdash\Code{reweight}(f,s):\stype \Code{S}\rightsquigarrow \gamma \mapsto \denote{f}(\gamma) \cdot \mu(\gamma)}
{\Gamma\vdash s:\stype \Code{ S} \rightsquigarrow \mu & \Gamma\vdash f:\Code{S \to R^+}}
}
\\ \\
\multicolumn{6}{c}{
\infer[\denote{f}: \denote{\Code{T}}\to\denote{\Code{T}}~\text{ergodic \wrt}~\mu, x~\text{typical}]
{\Gamma\vdash \Code{prng}(f,x)\rightsquigarrow\mu}
{}
}
\end{tabular}
\caption{Rules for asymptotic targeting}
\vspace{-4mm}
\label{table:targetingrules}
}
\end{table*}

\begin{figure*}[t]
{
\footnotesize
\centering
\[
{\infer{\vdash\Code{map(produce, reweight(accept, joint)) : \stype R^+} \rightsquigarrow \denote{\Code{produce}}_* (\denote{\Code{accept}} \cdot (U \otimes \mathrm{N}(0, 1))) = \Gamma(\alpha, 1)}
{\infer
{\vdash \Code{reweight(accept, joint) : \stype T} \rightsquigarrow \denote{\Code{accept}} \cdot (U \otimes \mathrm{N}(0, 1))}
{\infer{\Code{\vdash map(id_R \times box, rand^3) : \stype T} \rightsquigarrow \denote{\Code{id_R}\times\Code{box}}_* (U^3) = U \otimes \text{N(0, 1)}}{\vdash \Code{rand^3 : \stype S} \rightsquigarrow U^3 & \vdash \Code{id_R \times box : S \to T}} & \vdash \Code{accept : T \to R^+}} & \Code{\vdash produce : T \to R}
}
}
\]
}
\vspace{-4mm}
\caption{Validity of Marsaglia sampler in \cref{alg:marsagliamethod}}
\vspace{-4mm}
\label{fig:marsagliasamplerproof}
\end{figure*}

\begin{definition}\label{def:targeting}
We will say that the sampler \(\Gamma \vdash s: \Code{\stype S}\) \emph{asymptotically targets}, or simply \emph{targets}, the continuous map \(\mu: \denote{\Gamma} \to \mathcal{P} \denote{\Code{S}}\) if for every $\gamma\in\denote{\Gamma}$,
\[
\varepsilon_{\denote{\Code{S}}}\circ \denote{s}(\gamma)= \mu(\gamma).
\]
In particular, $\widehat{\denote{s}(\gamma)}_n$ always converges as $n\to\infty$; diverging samplers do not target anything.

\noindent We will say that \(\Gamma \vdash s: \stype \Code{S}\) is \emph{\(K\)-equidistributed} with respect to the morphism \(\mu : \denote{\Gamma} \to \mathcal{P}\denote{\Code{S}}\) if for every \(\gamma \in \denote{\Gamma}\), \(\varepsilon_{\denote{\Code{S}}} \circ \denote{s^K}(\gamma) = \mu^K(\gamma)\), where the self-product \(s^K\) is defined in \eqref{eq:selfproduct} and \(\mu^K(\gamma)\in\pr(\denote{\Code{S}}^K)\) is the \(K\)-fold product of the measure $\mu(\gamma)$ with itself.
\end{definition}

We introduce in \cref{table:targetingrules} a relation \(\rightsquigarrow\) which is sound with respect to asymptotic targeting; this is the relation which was used in the proofs of \cref{sec:examples}. That is, if \(\Gamma \vdash s: \stype \Code{S}\rightsquigarrow \mu\), then \(s\) is a parametrised sampler on \(\Code{S}\) which asymptotically targets a parametrised distribution \(\mu\) on \(\denote{\Code{S}}\). Here, we use Greek lower case letters $\mu,\nu$ to represent (parametrised) distributions in order to emphasise their role as meta-variables, used only in the context of the targeting calculus, and not within the language itself. In the rule for \(\Code{reweight}\), we abbreviate the operation of reweighting a measure \(\mu\) on \(X\) by \(f : X \to \Rnn\) as the product \((f \cdot \mu)(A) = \frac{\int_A f(x)\,d\mu(x)}{\int_X f(x)\,d\mu(x)}\), assuming that the integral in question is finite and nonzero.

\cref{table:targetingrules} incorporates a rule for building samplers from scratch as pseudo-random number generators defined by a deterministic endomap $t: \Code{T}\to\Code{T}$ and an initial value $x: \Code{T}$ via $\Code{prng}(t,x):\stype\Code{T}$.\footnote{Applying this rule requires showing that the initial point of the sampler is \emph{typical}; a point $x\in\denote{\Code{T}}$ is called \emph{typical} if it belongs to the $\mu$-mass 1 subset $X\subseteq \denote{\Code{T}}$ in which the ergodic theorem holds \cite[Theorem 9.6]{kallenberg1997foundations}.} However, \cref{table:targetingrules} also incorporates a set of `axioms' for built-in samplers \(\Code{rand}_i\), each targeting distributions $\mu_i\in\pr\denote{\Code{T}_i}$; in some settings, it may be defensible to assume access to `truly random' samplers which generate samples using a physical process.

The reader might wonder why \cref{table:targetingrules} does not have a rule for the \texttt{thin} operation: after all, if $\sigma$ is a sampler targeting a distribution $\mu$, then only keeping every $n$ samples should produce a good sampler as well.  Whilst this is true of the sequences produced by `true' i.i.d. samplers (for example physical samplers) with probability 1, this rule is in general not sound, as the following simple example shows.

\begin{example}
Consider the sampler on $\{0,1\}$ defined by the program $\Code{prng}(\la{x}{\Code{R}}{1-x},0)$.
This sampler, which generates the unweighted samples $(0,1,0,1,\ldots)$, targets the uniform Bernoulli distribution; however, applying $\mathtt{thin}(2,-)$ to it yields a sampler which targets the Dirac measure $\delta_0$.
\end{example}

This example highlights the fact that samplers can be manifestly non-random, and yet from the perspective of inference -- that is to say, from the perspective of the topology of weak convergence -- target bona fide probability distributions.

\begin{theorem}\label{thm:targeting_sound}\ifarxiv[\cref{proof:targeting_sound}]\else\cite[Appendix A]{ourarxivversion}\fi~Targeting \(\rightsquigarrow\) is sound: if \(\Gamma \vdash s:\Code{\stype S}\rightsquigarrow \mu\), then \(\varepsilon_{\denote{\Code{S}}} \circ \denote{s} = \mu\).
\end{theorem}


\ifarxiv It is easily seen that the natural corresponding notion of completeness, \(\varepsilon_{\denote{\Code{S}}} \circ \denote{\Code{s}}= \mu \to \Gamma \vdash s : \Code{\stype S} \rightsquigarrow \mu\), does not hold; see the following example.

\begin{example} Let \(s \defeq \Code{prng(\lambda n : R . n+1, 0)}\); clearly, \(s\) does not target any (probability) measure. Apply to \(s\) the transformation \(\Code{map}(\lambda \Code{n : R}~.~2^\wedge(\Code{-1*n}), s)\). While \(s\) does not target anything, and so our targeting calculus cannot prove that the transformed sampler targets anything, it is immediate that this sampler \emph{does} target the Dirac measure \(\delta_0\).\end{example}\fi

We saw in \cref{sec:empiricaltransformation} how our (sub-)typing system can be used to safely pushforward samplers through maps which are only piecewise continuous.  Our typing system also allows us to add additional constraints to samplers. Specifically,  we can ensure that a sampler visits certain subsets infinitely often.

\begin{proposition}\label{prop:targeting_subtypes}\ifarxiv[\cref{proof:targeting_subtypes}]\else\cite[Appendix A]{ourarxivversion}\fi~Assume \(\Gamma \vdash s : \Code{\stype S} \rightsquigarrow \mu\), \(\Code{S} \sub \Code{T}\) and $\denote{\Code{T}}$ second-countable;  then \(\Gamma \vdash \map(\lambda x : \Code{S}. \cast{T}{x}, s) : \stype\Code{T}\) targets the same measure \(\mu\) on \(\Code{T}\).  Moreover,  if $\denote{\Code{T}}$ is metrizable, if $U$ is in the topology of $\denote{\Code{S}}$ but not $\denote{\Code{T}}$ and $\mu(\partial_{\Code{T}} U)>0$ (where $\partial_{\Code{T}}$ denotes the boundary in $\denote{\Code{T}}$) then $s$ must visit $\partial_{\Code{T}} U$ \io (infinitely often).
\end{proposition}

\begin{example}
Suppose we want $s:\stype\Code{R}\rightsquigarrow \mathrm{Bern}(\nicefrac{1}{2})$.  A sampler alternating between the sampler $z \defeq \Code{prng}(\la{x}{\Code{R}}{x/2}, 1)$ of \cref{ex:discont0} and its shifted version $\Code{map}(\la{x}{\Code{R}}{1+x}, z)$ will satisfy the condition, but will never visit $0$ or $1$! We can use the previous result to enforce that a sampler $s$ targeting $\mathrm{Bern}(\nicefrac{1}{2})$ should visit $0$ \io by constructing $s$ in such a way that it has type $\stype\hspace{-3pt}\left((x,0)\inv\Code{Neq}+(x,0)\inv\Code{Eq}\right)$ (see \cref{ex:single_discont}). We can, in the same manner,  enforce that a sampler $s'$  targeting $\mathrm{Bern}(\nicefrac{1}{2})$ visits $1$ \io Finally,  using the last two rules of~\cref{lang:subtyping} which build the coarsest common refinement of two topologies, we can combine $s$ and $s'$ to create a sampler targeting $\mathrm{Bern}(\nicefrac{1}{2})$ and guaranteed to visit $0,1$ \io
\end{example}

\begin{example} We conclude with an example highlighting sampler compositionality by chaining two well-known sampling algorithms. Consider the following program:

\vspace{2mm}	
\begin{lstlisting}[caption=Marsaglia sampler for gamma random variables, label={alg:marsagliamethod}]
let box = $\lambda$u : R$^+\times$R$^+$ . sqrt(-2*log(fst(u))) * cos(2*pi*snd(u)) in
let joint = map(id$_R$$\times$box, rand$^3$) in
let d = $\alpha$ - 1/3 in
let c = 1/(3*sqrt(d)) in
let accept = $\lambda$(u, x) : T .
	let v = (1+c*x)^3 in
		if v > 0 and log(u) < x^2 + d - d*v + d*log(v) then 1 else 0 in
let produce = $\lambda$z : T . d*(1+c*fst(cast$\langle$R$\times$R$\rangle$z))^3 in 
	map(proj, reweight(accept, joint))
\end{lstlisting}
First, the Box-Muller technique is a well-known technique for generating standard normal random variates using two independent uniform samples; its verification using the $\Code{map}$ rule of \cref{table:targetingrules} is straightforward. We then form a joint sampler consisting of independent uniform and Gaussian samples, and then consume both of these samples to generate gamma-distributed random variables \(z \sim \Gamma(\alpha, 1)\), for shape \(\alpha \geq 1\), using a well-known rejection sampling technique; see \cite{marsaglia2000gamma} for a proof. The validity of this sampler is sketched in \cref{fig:marsagliasamplerproof}; we omit some types for brevity.
\end{example}

\section{Discussion}

We have presented a `probabilistic' language designed to compositionally construct \emph{samplers}. We have given this language an intuitive operational semantics and a denotational semantics in the category of CG-spaces, and shown that the two are equivalent for closed samplers. This denotational universe is sufficiently rich to interpret sampler types coinductively, and to interpret functions which are only piecewise-continuous on the standard topologies given by the type system (\cref{sec:denotationalsemantics}).

With the support of this language, we have shown how to compositionally reason about the validity of sampler constructions, an essential aspect in the practice of probabilistic programming. Our approach draws on a sound equational system to reason about equivalent ways of constructing the same sampler (\cref{sec:equivalence}) and a sound system for reasoning about semantic correctness (\cref{sec:verification}).

What distinguishes our approach is that we are in effect providing a purely deterministic semantics for probabilistic programs. This approach is much closer to the practice of probabilistic programming, in which samples and samplers are the most important concrete entities; this distinction between samplers and the measures they target is necessary in order to support pseudo-random number generation. Measure-theoretic entities, which have typically been a part of the denotational semantics of probabilistic languages, \eg \cite{K81c,heunen2017convenient,ehrhard2017measurable,vakar2019domain,dahlqvist2019semantics}, instead take a meta-theoretic role as verification criteria.

Two commonly-used schemes for producing samplers are missing from our calculus: Markov chain Monte Carlo methods and resampling techniques, as applied in \eg particle filters. We consider adding these constructions to our language, together with the corresponding correctness proofs in our targeting calculus, to be future work.

\newpage

\bibliographystyle{acm}
\bibliography{bibliography}

\ifarxiv
\newpage

\newpage
\onecolumn

\appendix

\subsection*{Type-checking of Rejection-sampling~(\cref{sec:examplerejectionsampling})}
We first type-check \(\Code{accept}\) from~\cref{sec:examplerejectionsampling}. This determines the type $\Code{T}$ which was left unspecified. To keep the derivation readable we define $t\defeq (y,\Code{phi}(x)*\Code{sqrt}(2*\Code{pi}))$.
\begin{figure*}[h!]
	\[
	\infer{\vdash \lambda (x,y): t\inv(\predinv{<}{0})+t\inv(\predinv{<}{1})~.~\ite{y<\Code{phi}(x)*\Code{sqrt}(2*\Code{pi})}{1}{0}:t\inv(\predinv{<}{0})+t\inv(\predinv{<}{1})\to \Code{R^+}}
	{
		\infer{(x,y): t\inv(\predinv{<}{0})+t\inv(\predinv{<}{1}) 			\vdash\ite{y<\Code{phi}(x)*\Code{sqrt}(2*\Code{pi})}{1}{0}:\Code{R^+}}
		{
			\infer{(x,y): t\inv(\predinv{<}{0})+t\inv(\predinv{<}{1}) \vdash y<\Code{phi}(x)*\Code{sqrt}(2*\Code{pi}):\Code{B} \quad \vdash 0:\Code{R^+} \quad \vdash 1:\Code{R^+}}
			{
				\infer[\predinv{<}{0}+\predinv{<}{1}\sub \Code{R\times R}]
				{(x,y): t\inv(\predinv{<}{0})+t\inv(\predinv{<}{1}) \vdash (y,\Code{phi}(x)*\Code{sqrt}(2*\Code{pi})): \predinv{<}{0}+\predinv{<}{1}}
				{
					\infer{x:\Code{R}, y:\Code{R}\vdash (y,\Code{phi}(x)*\Code{sqrt}(2*\Code{pi})): \Code{R\times R}}
					{
						x:\Code{R}, y:\Code{R}\vdash y:\Code{R} & \infer{x:\Code{R}, y:\Code{R}\vdash \Code{phi}(x)*2*\Code{pi}:\Code{R}}
						{
							x:\Code{R}, y:\Code{R}\vdash x:\Code{R} \quad \vdash\Code{phi:R\to R} \quad\vdash 2:\Code{R} \quad \vdash \Code{pi}:\Code{R}
						}
					}
				}
			}
		}
	}
	\]
	\caption{Derivation of \(\Code{accept}\) from \cref{sec:examplerejectionsampling}}
	\label{fig:typecheckingrejection1}
\end{figure*}

We now define $\Code{T}=t\inv(\predinv{<}{0})+t\inv(\predinv{<}{1})$ and $\Code{proj}\defeq \la{u}{\Code{T}}{\Code{fst(cast}\langle\Code{R\times R}\rangle u)}$.

\begin{figure*}[h!]
	\[
	\infer{\vdash \map(\Code{reweight(accept, tri\otimes rand),proj}):\stype\Code{R}}
	{
		\infer{\vdash\Code{proj:T\to R}}	
		{
			\infer{u:\Code{T}\vdash \Code{fst(cast}\langle\Code{R\times R}\rangle u):\Code{R\times R}}
			{
				\infer[\Code{T\sub R\times R}]{	
					u:\Code{T}\vdash \Code{cast}\langle\Code{R\times R}\rangle u:\Code{R\times R}}{u:\Code{T}\vdash u:\Code{T}}
			}
		}
		&
		\infer{\vdash \Code{reweight(accept, tri\otimes rand)}):\stype\Code{T} }
		{\vdash \Code{accept: T\to R^+} & 
			\infer[\stype \Code{T}\sub \stype (\Code{R\times R}), \Gamma=\emptyset]{\vdash \Code{tri\otimes rand:\stype T}}
			{
				\infer{\Code{\vdash tri\otimes rand}:\stype (\Code{R\times R})}
				{\vdash \Code{tri}:\stype\Code{R}\quad\vdash \Code{rand}:\stype\Code{R}
				}
			}
		}
	}
	\]
	\caption{Derivation of the rejection sampling algorithm~\cref{sec:examplerejectionsampling}}
	\label{fig:typecheckingrejection2}
\end{figure*}

\begin{table*}[t]
{
\centering
\footnotesize
\makebox[\textwidth]{%
\begin{tabular}{c c c}
\infer[v~\text{a value}]
{(v,N) \to v}
{}
&
&
\infer[\mathrm{Func}\ni f: \Code{T}\to \Code{G}]{(f(t),N)\to f(v)}{(t, N)\to v}
\\ \\
\infer{(\letx{x}{s}{t}, N) \to v}{((\lambda x:T.t)(s), N) \to v} 
& & 
\infer{((\lambda x:\Code{T}.t)(s), N) \to v}{(t[x \leftarrow s], N) \to v} \\ \\
\infer{(\cast{T}{t}, N) \to v}{(t, N) \to v}
&
\infer{(\case{(c, t)}{x}{s}{I}{i}), N) \to v}{(c, N) \to j\in I & (s_j[x_j \leftarrow t], N) \to v}
&
\infer{(\ini{j}{t}, N) \to v}{(t, N) \to v}
\\ \\
\infer{((s,t), N)\to (v_1,v_2)}
{(s,N)\to v_1 & (t,N)\to v_2}
&
\infer{(\Code{fst}(t), N) \to v_1}{(t, N) \to (v_1, v_2)} 
& 
\infer{(\Code{snd}(t), N) \to v_2}{(t, N) \to (v_1, v_2)}
\\ \\
\multicolumn{3}{c}{
\infer{(\Code{map}(s, t), N) \to ((v_1, w_1), \ldots, (v_{N}, w_{N}))}
{\left((s(\hd(t)),\wt(t)), N\right) \to (v_1,w_1) &  \ldots & ((s(\hd(\tl^{N-1}(t)),\wt(\tl^{N-1}(t))), N) \to (v_{N},w_N)}
} 
\\ \\
\multicolumn{3}{c}{
\infer{(\Code{reweight}(s, t), N) \to ((v_1, w_1 ), \ldots, (v_{N}, w_{N})}
{((\hd(t), s(\hd(t))\cdot \wt(t)),N)\to(v_1,w_1)&\ldots & ((\hd(\tl^{N-1}(t)), s(\hd(\tl^{N-1}(t)))\cdot \wt(\tl^{N-1}(t))),N)\to(v_N,w_N)}
} 
\\ \\
\multicolumn{3}{c}{
\infer{(s \otimes t, N) \to (((v_1, v_1'), w_1 \cdot w_1'), \ldots, ((v_{N}, v_{N}'), w_{N} \cdot w_{N}'))}{(s, N) \to ((v_1, w_1), \ldots, (v_{N}, w_{N})) & (t, N) \to ((v_1', w_1'), \ldots, (v_{N}', w_{N}'))}
} 
\\ \\
\infer{(\Code{hd}(t), N) \to v_1}{(t, N) \to ((v_1, w_1), \ldots, (v_{N}, w_{N}))}
& 
\infer{(\Code{tl}(t), N-1) \to ((v_2, w_2), \ldots, (v_N, w_N))}{(t, N) \to ((v_1, w_1), \ldots, (v_N, w_N))} 
& 
\infer{(\Code{wt}(t), N) \to w_1}{(t, N) \to ((v_1, w_1), \ldots, (v_N, w_N))}
\\ \\
\multicolumn{3}{c}{
\infer{(\Code{thin}(s, t), N) \to ((v_1, w_1), (v_{i+1}, w_{i+1}), (v_{2i+1}, w_{2i+1}), \ldots, (v_{(Ni)+1}, w_{(Ni)+1}))}{(s, N) \to i & (t, Ni) \to ((v_1, w_1), \ldots, (v_{Ni}, w_{Ni}))}
}
\\ \\
\multicolumn{3}{c}{
\infer{(\Code{prng}(s, t), N) \to ((v_1, 1), \ldots, (v_{N}, 1))}
{(t, N) \to v_1 & (s(t), N) \to v_2 & \ldots & (s^{N-1}(t), N) \to v_{N}}
}
\end{tabular}
}}
\caption{Full big-step operational semantics}
\label{table:operationalfull}
\end{table*}

\begin{appendixproof}[\cref{prop:opsemvalues}]\label{proof:opsemvalues}
In order to prove this result by induction on the derivation tree of \((t, N) \to v\), we must first generalise it to include higher samplers.
\begin{proposition}
If \(\vdash s : \stype^k \Code{S}\) is a closed \(k\)-order sampler where \(\Code{S}\) is not a sampler type and \(k \in \{0, 1, 2, \ldots\}\), then for any \(N \in \N\), if \((s, N) \to v\), then \(v\) has the form of a \(k\)-nested weighted list of values of type \(\Code{S}\). For example, for \(k=0\), \(v : \Code{S}\) is simply a value of type \(\Code{S}\); for \(k=1\), \(v = ((v_1, w_1), \ldots, (v_N, w_N))\) is a weighted list of values \(v_n : \Code{S}\) and \(w_n \geq 0\); for \(k=2\), \(v = (((v_1^1, w_1^1), \ldots, (v_N^1, w_N^1)), w_1), \ldots, (((v_1^N, w_1^N), \ldots, (v_N^N, w_N^N)), w_N))\) is a weighted list of weighted lists of values of type \(v_n^n : \Code{S}\), and so on.
\end{proposition}

\textbf{Base case.} As values \(v\) cannot have sampler type, the only possibility for a derivation \((v, N) \to v\) where \(v : \stype^k \Code{T}\) for some type \(\Code{T}\) is \(k=0\), which makes our result immediate.

\textbf{Inductive case.} We illustrate the inductive argument for each case, depending on the last rule of the derivation of \((t, N) \to v\), where \(\vdash t : \stype^{k} \Code{T}\) is a \(k\)-order sampler for some \(k \in \{0, 1, 2, \ldots\}\), and \(\Code{T}\) is not a sampling type (i.e. contains no occurrences of \(\stype\)).

\begin{enumerate}
\item \textbf{Built-in functions.} There are no built-in functions which either input or output sampler types, so \(k=0\). Taking the inductive hypothesis that each input \(s_i : \Code{G}_i\) reduces to a value \(v_i : \Code{G}_i\), where \(\Code{G}_n\) are ground types, we immediately obtain that \((f(s_1, \ldots, s_n), N) \to v\) evaluates to a value \(v\) of ground type \(\Code{G}\), giving our result.
\item \textbf{Case.} Assuming that \(t = \Code{case}~(c, t')~\Code{of}~\{(i, x_i) \Rightarrow s_i\}_{i \in n}\), we must have \(\vdash s_i : \stype^{k} \Code{T}\). Taking the inductive hypothesis that \((s_i[x_i \leftarrow t'], N) \to v\) evaluates to a \(k\)-nested weighted list, if \((c, N) \to i \in n\), it immediately follows that \((t, N) \to v\) does as well.
\item \textbf{Function application.} Take \(t = (\lambda x : \Code{\stype^{k'} S} . t')(s)\) to be an instance of function application, where the function in question inputs a \(k'\)-sampler and outputs a \(k\)-sampler, where \(\Code{S}\) does not contain any sampler types itself; we must have \(\vdash s : \stype^{k'} \Code{S}\) in order for the expression to be well-typed. In order to have \((t, N) \to v\) evaluate to a value, our operational semantics requires \((t'[x \leftarrow s], N) \to v\); taking the inductive hypothesis that \(t'[x \leftarrow s]\) evaluates to a \(k\)-nested list of values of type \(\Code{S}\), our desired result follows.
\item \textbf{\(\Code{let}\)-binding.} Trivially follows from function application, as \((\Code{let}~x=s~\Code{in}~t', N) \to v\) iff \(((\lambda x:S.t')(s), N) \to v\).
\item \textbf{Product.} If \(t = (s, s')\), the result is trivial as \(\vdash t : \stype^{k} \Code{T}\) implies \(k=0\) and so \(\Code{T} = \Code{S} \times \Code{S'}\) where \(\vdash s : \Code{S}, \vdash s' : \Code{S'}\) are each not sampler types; therefore, \((s, s')\) is clearly a value of type \(\Code{T}\) (\ie, a 0-nested weighted list).
\item \textbf{Projections.} If \(t = \Code{fst}((s, s'))\), then \(\vdash s : \stype^{k} \Code{T}\), and so the inductive hypothesis \((s, N) \to v\) immediately implies our result; the same argument applies to \(\Code{snd}\) and \(t\).
\item \textbf{Head.} If \(t = \Code{hd}(s)\), then \(\vdash s : \stype^{k+1} \Code{T}\). Taking the inductive hypothesis that \((s, N) \to v\) implies that \(v\) is a \((k+1)\)-nested weighted list of values of type \(\Code{T}\), we need only note that the first element of this list is itself a \(k\)-nested weighted list of values of type \(\Code{T}\).
\item \textbf{Weight.} If \(t = \Code{wt}(s)\), our result is trivially true, as the output of \(\Code{wt}(s)\) can only be a nonnegative real number \(\vdash t : R^+\).
\item \textbf{Tail.} If \(t = \Code{tl}(s)\), then by our inductive hypothesis, \((s, N+1) \to ((v_1, w_1), \ldots, (v_{N+1}, w_{N+1}))\) where each \(v_n\) is a \((k-1)\)-nested weighted list of elements of type \(\Code{T}\). It immediately follows that \((\Code{tl}(s), N)\) evaluates to a weighted list of \(N\) elements whose elements are each \((k-1)\)-nested lists of type \(\Code{T}\).
\item \textbf{Thin.} If \(t = \Code{thin}(i, s)\), then \(\vdash n : N\) and \(\vdash s : \stype^{k} \Code{T}\), and by our inductive hypothesis, \((s, Ni) \to ((v_1, w_1), \ldots, (v_{Ni+1}, w_{Ni+1}))\) where each \(v_n\) is a \((k-1)\)-nested weighted list of elements of type \(\Code{T}\). It immediately follows that \((\Code{thin}(i, s), N)\) evaluates to a weighted list of \(N\) elements whose elements are each \((k-1)\)-nested lists of type \(\Code{T}\).
\item \textbf{Map.} If \(t = \Code{map}(s, t')\) and \((t, N) \to v\), the operational semantics of \(\Code{map}\) requires that 
\[
\left((s(\Code{hd}(\Code{tl}^{n-1}(t'))), \Code{wt}(\Code{tl}^{n-1}(t'))), N\right) \to(v_n, w_n)
\]
for each \(n \in \{1, \ldots, N\}\). As \(t'\) is a subterm of \(t\), our inductive hypothesis implies that if \(\vdash t' : \stype^{k'} \Code{S}\) for some \(k' \in \{1, 2, \ldots\}\), then for any \(N \in \N\), \(t'\) evaluates to a \(k'\)-nested weighted list of values of type \(\Code{S}\).  Note that in order for \(t\) to be well-typed, we must have \(\vdash s : \stype^{k'-1} \Code{S} \to \stype^{k-1} \Code{T}\). We have already shown that if this is the case, then \(\Code{tl}^{n-1}(t')\) evaluates to a \(k'\)-nested weighted list of values of type \(\Code{S}\), and then that \(\Code{hd}(\Code{tl}^{n-1}(t'))\) evaluates to a \((k'-1)\)-nested weighted list of values of type \(\Code{S}\) of length \(N\), and then that \(s(\Code{hd}(\Code{tl}^{n-1}(t')))\) evaluates to a \((k-1)\)-nested weighted list of values of type \(\Code{T}\) of length \(N\). Our result follows by observing that if \(((s(\Code{hd}(\Code{tl}^{n-1}(t'))), \Code{wt}(\Code{tl}^{n-1}(t'))), N) \to (v_n, w_n)\) for each \(n \in \{1, \ldots, N\}\) where each \(v_n\) is a \(k-1\)-nested weighted list of values of type \(\Code{T}\), then the expression \(((v_1, w_1), \ldots, (v_N, w_N))\) is a \(k\)-nested weighted list of values of type \(\Code{T}\), completing the proof.
\item \textbf{Reweight.} This proof works in exactly the same way as that of \(\Code{map}\).
\item \textbf{Product of samplers.} If \(t = s \otimes s'\) and \(\vdash t : \stype^{k} \Code{T}\) where \(\Code{T}\) contains no instances of \(\stype\), then it must be that \(k \geq 1\), that \(\vdash s : \stype^{k} \Code{S}\), and that \(\vdash s' : \stype^{k} \Code{S'}\). Our inductive hypothesis states that \((s, N) \to ((v_1, w_1), \ldots, (v_N, w_N))\) where each \(v_1\) is a \((k-1)\)-nested weighted list of values of type \(\Code{S}\), and \((s', N) \to ((v_1', w_1'), \ldots, (v_N', w_N'))\) where each \(v_1'\) is a \((k-1)\)-nested weighted list of values of type \(\Code{S'}\). Our result then follows by noting that the product \((((v_1, v_1'), w_1 \cdot w_1'), \ldots, ((v_N, v_N'), w_N'))\) is a \(k\)-nested weighted list of values of type \(\Code{S}\).
\item \textbf{Pseudorandom number generators.} Finally, assume \(t = \Code{prng}(s, t')\); in order for this expression to be well-typed, we must have \(k \geq 1\), \(\vdash t' : \stype^{k-1} \Code{T}\), and \(\vdash s : \stype^{k-1} \Code{T} \to \stype^{k-1} \Code{T}\). In order for \((t, N)\) to evaluate to anything, we must have \((s^{n-1}(t), N) \to v_n\) for each \(n \in \{2, \ldots, N\}\); as we have already proven the case for function abstraction, we know that each \(v_n\) is a \((k-1)\)-nested weighted list of values of type \(\Code{T}\). We need only note then that \(((v_1, 1), \ldots, (v_N, 1))\) is clearly a \(k\)-nested weighted list of values of type \(\Code{T}\).
\end{enumerate}
\end{appendixproof}

\begin{appendixproof}[\cref{prop:beh}]\label{proof:continuousbeh}
Let $f_n\to f$ be a convergent sequence a coalgebra maps in $[X,FX]$; we need to show that $\beh_X(f_n)\to \beh_X(f)$ in $[X,\nu F]$. The topology on $[X,\nu F]$ is the compact-open topology, which means that it is generated by the subbase of open sets of the shape 
\[
(K,V)\defeq \{h: X\to\nu F\mid h[K]\subset U\}
\]
for some fixed compact set $K\subseteq X$ and open set $U\subseteq \nu F$. Moreover, by construction of $\nu F$ (see \cref{thm:adamek}), we know that the topology is induced by the product topology on $\prod_i F^i 1$. A base for this topology is given by intersections of cylinder sets with $\nu F$. 
Because we are also assuming that $\interior{\nu F}\neq \emptyset$ in $\prod_i F^i 1$,  it contains such an open set, and we can thus simply start with an open neighbourhood of $\beh_X f$ of the shape $(K,\prod_i V_i)$ where for all but \emph{finitely} many indices $V_i=F^i 1$, and for the other indices $V_i$ is an open subset of $F^i1$ (and we don't have to worry about intersecting with $\nu F$). Given such an open set, we need to find $N\in\N$ such that for all $n>N$ $\beh_X(f_n)\in(K, \prod_i V_i)$.

By the construction of \cref{thm:adamek} we have that
\[
\beh_X(f)(x) = (!_X(x), F!_X(f(x)), F^2!_X(Ff(f(x))),\ldots)
\]
where $!_X: X\to 1$ is the unique morphism to the terminal object. For each of the finitely many non-trivial open subsets $V_{i_k}\subset F^{i_k}1, 1\leq k\leq M$, because $f_n\to f$ and composition with continuous functions is a continuous operation on internal hom sets in $\CG$ (\cite[5.9]{steenrod1967convenient}), it follows that there exists $N_k$ such that for every $n>N_k$
\[
F^{i_k}!_X \circ F^{i_k-1}f_n\circ \ldots\circ f_n\in (K,V_{i_k})
\] 
By taking $N=\max_{1\leq k\leq M}N_k$, we get that for for \emph{all} $i\in \N$ and all $n>N$ 
\[
F^{i}!_X \circ F^{i-1}f_n\circ \ldots\circ f_n\in (K,V_{i})
\]
In other words, for any $n>N$, $\beh_X(f_n)\in (K,\prod_i V_i)$, which concludes the proof.
\end{appendixproof}

\begin{appendixproof}[\cref{thm:adequacy}]\label{proof:adequacy}
\noindent $\Leftarrow$) By induction on the derivation tree of $(t,N)\to v$. 

\textbf{Base case.}  The base case is trivial: the only derivation of length 0 allowed by  \cref{table:operational} assumes that $t=v$ is a value. It is easy to check that if $t:\Code{T}$ is a value, then $\pnt=\id_{\denote{\Code{T}}}$ and thus $\pnt(\denote{t})=\denote{t}=\denote{v}$ tautologically.

\textbf{Inductive case.}  Assume that the last rule of the derivation of $(t,N)\to v$ is
\begin{enumerate}[(i)]
\item \textbf{Built-in functions.} $t=f(s_1,\ldots,s_n)$ for some $s_i:\Code{G}_i, 1\leq i\leq n$.  Since for a ground type $\Code{G}$ we have $\pnt[G]=\id_{\denote{\Code{G}}}$, we immediately get
\begin{align*}
\pnt[G](\denote{f(s_1,\ldots,s_n)})&\defeq \denote{f}(\denote{s_1}, \ldots\denote{s_n})\\
&=\denote{f}(\denote{v_1},\ldots,\denote{v_n}) & \text{induction hypothesis}
\end{align*}
\item \textbf{Case.} $t=\case{(c, t')}{x}{s}{I}{i}$. If \(c\) chooses the branch \(j \in n\), then
\begin{align*}
\pnt\left(\denote{\case{(c, t')}{x}{s}{I}{i}}\right)&\defeq \denote{s_j}(\denote{t'}) \\
&=\denote{v}&\text{induction hypothesis}
\end{align*}
\item $\lambda$\textbf{-abstraction.} $t=(\lambda x:\Code{S}.~t')(s):\Code{T}$ for some $s:\Code{S}$ and some $t':\Code{T}$.
\begin{align*}
\pnt\left(\denote{(\lambda x:\Code{T}.~t')(s)}\right)&\defeq \pnt\circ \ev_{\denote{\Code{S}},\denote{\Code{T}}}\left(\denote{\lambda x:\Code{T}.~t'}\times{\denote{s}}\right)
\\
& \defeq \pnt\circ \ev_{\denote{\Code{S}},\denote{\Code{T}}}
\left(\widehat{\denote{t'}}\times\denote{s}\right)
& \text{Currying }\denote{t'}
\\
&=\pnt\circ \ev_{\denote{\Code{S}},\denote{\Code{T}}}
\left(\denote{t'}\times\denote{s}\right) & \denote{t'}\text{ has only one variable}
\\
&=\pnt\left(\denote{t'}(\denote{s})\right)
\\
&=\denote{v} & \text{induction hypothesis}
\end{align*}
\item \texttt{let}\textbf{-binding.} $t=\letx{x}{s}{t'}:\Code{T}$ for some $s:\Code{S}$ and $t':\Code{T}$.
\begin{align*}
\pnt\left(\denote{\letx{x}{s}{t}}\right)&\defeq \denote{t'}(\denote{s})
\\
&=\pnt\left(\denote{\lambda x.~t'}(\denote{s})\right)
\\
&=\denote{v} & \text{induction hypothesis}
\end{align*}
\item \textbf{Product.} $t=(s,s')$ for some $s: \Code{S}, s':\Code{S'}$
\begin{align*}
\pnt[S\times S'](\denote{(s,s')})&\defeq \pnt[S\times S']\left(\langle\denote{s},\denote{s'}\rangle\right)
\\
&=\langle\pnt[S](\denote{s)},\pnt[S'](\denote{s'})\rangle & \text{inductive definition of }\pnt\\
&=\langle\denote{v_1},\denote{v_2}\rangle & \text{induction hypothesis}
\\
&=\denote{(v_1,v_2)}
\end{align*}
\item \textbf{Projections} $t=\fst(s,s')$ for some $s: \Code{S}, s':\Code{S'}$ 
\begin{align*}
\pnt[S]\left(\denote{\fst(s,s')}\right)&\defeq \pnt[S]\left(\pi_1 \langle \denote{s},\denote{s'}\rangle\right)
\\
&= \pnt[S](\denote{s})\\
&=\denote{v_1}& \text{induction hypothesis}
\end{align*}
and similarly for $\snd$.
\item \textbf{Pushforward.} $t=\map(s,t)$ for some $s: \Code{S}\to\Code{T}$ and $t:\stype\Code{S}$.  To keep the derivation readable we will write $s$ instead of $\denote{s}$, $t$  instead of $\denote{t}$ and we introduce the following notation. Let $F$ denote the functor $\denote{\Code{T}}\times\R^+\times\Id$, let $\gamma: \nu F\to F\nu F$ denote the terminal coalgebra structure map $\mathrm{unfold}_{\Code{T}}$, let $\delta\defeq \denote{s}\times\id_{\R^+}\times \id_{\stype\Code{S}}\circ \mathrm{unfold}_{\Code{S}}$, the coalgebra structure map defining the \texttt{map} operation,  let $b=\beh(\delta)$, and let 
\begin{align*}
&h\defeq \pi_1\circ \mathrm{unfold}_{\Code{S}}  &\text{\ie  $h(t)$ is the first sample of $t$} \\
&w\defeq \pi_2\circ \mathrm{unfold}_{\Code{S}}&\text{\ie  $w(t)$ is the weight of the first sample of $t$} \\
&f\defeq \pi_3\circ \mathrm{unfold}_{\Code{S}}&\text{\ie  $f(t)$ is the tail of $t$} 
\end{align*}
With this we can now derive
\begin{align*}
&\hspace{-2em}\pnt[\stype T](\denote{\map(s,t)})\\
&\hspace{-2em}\defeq \pi_{1:N}\circ \left(\pnt[T\times R^+]\right)^\omega(\denote{\map(s,t)})
\\
&\hspace{-2em}\defeq \pi_{1:N}\circ \left(\pnt[T\times R^+]\right)^\omega(b(t))
\\
&\hspace{-2em}\stackrel{(1)}{=} \left(\pnt[T\times R^+]\right)^N\circ \pi_{1:N}(b(t))
\\
&\hspace{-2em}\stackrel{(2)}{=}\left(\pnt[T\times R^+]\right)^N \circ F\pi_{1:N-1}\circ \gamma\circ(b(t))
\\
&\hspace{-2em}\stackrel{(3)}{=}\left(\pnt[T\times R^+]\right)^N\circ F^{N-1}\pi_1\circ F^{N-2}\gamma\circ \ldots\circ F^0\gamma(b(t))
\\
&\hspace{-2em}\stackrel{(4)}{=}\left(\pnt[T\times R^+]\right)^N\circ F^{N-1}\pi_1\circ F^{N-1}b\circ F^{N-2}\delta\circ \ldots\circ F^0\delta(t) 
\\
&\hspace{-2em}\stackrel{(5)}{=}\left(\pnt[T\times R^+]\right)^N\circ F^{N-1}\pi_1\circ F^{N-1}b\left((s(h(t)),w(t)),\ldots,(s(h(f^{N-1}(t))),w(f^{N-1}(t))), f^{N-1}(t)\right)
\\
&\hspace{-2em}\stackrel{(6)}{=}\left(\pnt[T\times R^+]\right)^N\circ \left((s(h(t))),w(t)),\ldots,(s(h(f^{N-1}(t))),w(f^{N-1}(t)))\right)
\\
&\hspace{-2em}=\left(\pnt[T\times R^+](s(h(t))),w(t)),\ldots,\pnt[T\times R^+](s(h(f^{N-1}(t))),w(f^{N-1}(t)))\right)
\\
&\hspace{-2em}=\left((\pnt(s(h(t))),w(t)),\ldots,(\pnt(s(h(f^{N-1}(t)))),w(f^{N-1}(t)))\right)
\\
&\hspace{-2em}\stackrel{(7)}{=} \denote{((v_1,w_1), \ldots,(v_N,w_N))}
\end{align*}
where $(1)$ is the simple observation that $\pi_{1:N}\circ (\pnt)^\omega=(\pnt)^N\circ \pi_{1:N}$,  $(2)$ is by definition of $\gamma$, $(3)$ is by iteration of $(2)$, $(4)$ follows from the fact that $b$ is a coalgebra morphism, $(5)$ is by definition of $\delta$,  $(6)$ is by definition of $F$, $b$ and $p^1_{\Code{T}}$, and $(7)$ is by the induction hypothesis on the $N$ premises of the rule.
\item \textbf{Reweight.} The proof is very similar to the case of $\mathtt{map}$. Again, writing 
\[
\delta\defeq \id_{\denote{\Code{T}}}\times (-\cdot-)\times \id_{\denote{\stype\Code{T}}}\circ \langle\id_{\denote{\Code{T}}},\denote{s}\rangle\times \id_{\R^+}\times \id_{\denote{\stype\Code{T}}}\circ \gamma
\]
for the coalgebra structure defining \texttt{reweight} and $b=\beh(\delta)$, we get
\begin{align*}
&~\pnt[\stype T](\rew(s,t))\\
& \defeq \pi_{1:N}\circ \left(\pnt[T\times R^+]\right)^\omega(\rew(s,t))\\
& \defeq \left(\pnt[T\times R^+]\right)^N\circ \pi_{1:N}\circ b(t)\\
& \stackrel{(1)}{=}\left(\pnt[T\times R^+]\right)^N\circ F^{N-1}\pi_1\circ F^{N-1}b\left((h(t),s(h(t))w(t)), \ldots, (h(f^{N-1}(t)), s(f^{N-1}(t))w(f^{N-1}(t))),f^{N-1}(t)\right)\\
& \stackrel{(2)}{=}\left((\pnt(h(t)),s(h(t))w(t)), \ldots, (\pnt(h(f^{N-1}(t))), s(f^{N-1}(t))w(f^{N-1}(t)))\right)\\
&  \stackrel{(3)}{=} \left((\denote{v_1}, \denote{w_1}), \ldots,(\denote{v_N}, \denote{w_N})\right)
\end{align*}
where $(1)$ follows the same derivation as in the case of \texttt{map} but with the definition of $\delta$ as above,  $(2)$ is by definition of $F$ and $\pnt[T\times R^+]$, and $(3)$ is by the the induction hypothesis applied to the $N$ premises of the $\rew$ rule.
\item \textbf{Product of samplers.} The proof works in exactly the same way as for $\map$ and $\rew$.
\item \textbf{Thin.} The proof works in exactly the same way as for $\map$ and $\rew$.
\item \textbf{Pseudorandom number generators.} Consider the term $\prng(s,t):\stype\Code{T}$. Using
\[
\delta\defeq \langle\id_{\denote{\Code{T}}},1,\denote{s}\rangle
\]
and $b=\beh(\delta)$, the same steps as in the case of $\map$ and $\rew$ yield
\begin{align*}
\pnt[\stype T](\denote{\prng(s,t)})&\defeq \pi_{1:N}\circ (\pnt[T\times R^+])^\omega(\denote{\prng(s,t})\\
&=\left(\pnt[T\times R^+]\right)^N\circ F^{N-1}\pi_1\circ F^{N-1}b((t,1),(s(t),1),\ldots,s^{N-1}(t),1), s^{N}(t))\\
&=((\pnt(t),1),(\pnt(s(t)),1),\ldots,(\pnt(s^{N-1}(t)),1))\\
&=\denote{(v_1,1),(v_2,1),\ldots,(v_N,1))}
\end{align*}
\item \textbf{Head.} Consider the term $\hd(t)$ for some $t:\stype\Code{T}$.  Using the same notation as above
\begin{align*}
\pnt \circ \denote{\hd(t)}&\defeq \pnt\circ  \pi_1\circ \gamma(\denote{t})\\
&= \pi_1\circ \pi_1 \circ \left(\pnt[T\times R^+]\right)^N\circ \pi_{1:N}(\denote{t})\\
& =\pi_1\circ \pi_1 \denote{(v_1,w_1),\ldots,(v_N,w_N)} &\text{induction hypothesis} \\
&= \denote{v_1}
\end{align*}
\item \textbf{Weight.} Consider the term $\wt(t)$ for some $t:\stype\Code{T}$. The proof is the same as the above:
\begin{align*}
\pnt \circ \denote{\wt(t)}&\defeq \pnt\circ  \pi_2\circ \gamma(\denote{t})\\
&= \pi_2\circ \pi_1 \circ \left(\pnt[T\times R^+]\right)^N\circ \pi_{1:N}(\denote{t})\\
& =\pi_2\circ \pi_1 \denote{(v_1,w_1),\ldots,(v_N,w_N)} &\text{induction hypothesis} \\
&= \denote{w_1}
\end{align*}
\item \textbf{Tail.} Consider the term $\tl(t)$ f for some $t:\stype\Code{T}$. It is immediate that
\begin{align*}
\pnt[\stype T]\circ\denote{\tl(t)}&\defeq  \left(\pnt[T\times R^+]\right)^N \circ \pi_{1:N}(\pi_3\circ \gamma(\denote{t}))\\
&=\left(\pnt[T\times R^+]\right)^N \circ \pi_{2:N+1}(\denote{t})\\
&=\pi_{2:N+1}\circ \left(\pnt[T\times R^+]\right)^{N+1}\circ \pi_{1:N+1}(\denote{t})\\
&=\pi_{2:N+1} \denote{(v_1,w_1),\ldots,(v_{N+1},w_{N+1})} &\text{inductive hypothesis}\\
&=\denote{(v_2,w_2),\ldots,(v_{N+1},w_{N+1})}
\end{align*}
\end{enumerate}

\noindent $\Rightarrow$) By induction on the typing-proof of $t$. Note that for any term $t:\Code{T}$, $\pnt\denote{t}$ is necessarily a value, by definition of $\pnt$.

\textbf{Base case.} The only programs which are type-checkable in 0 steps are the constants. Since all constants are values and values operationally evaluate to themselves, the base case holds trivially.

\textbf{Inductive case.} The proof is routine and we only show a few cases.  Suppose that the last step of the rule applied in the type-checking of $t$ was 
\begin{enumerate}[(i)]
 \item \textbf{Product.} Suppose $\vdash (s,t):\Code{S\times T}$ and that $\pnt[S\times T](\denote{(s,t)})=\denote{v}$ for some value $v$. Since the last applied rule had premises $\vdash s:\Code{S}$ and $\vdash t:\Code{T}$ we have
\begin{align*}
 \denote{v}&=\pnt[S\times T](\denote{s\otimes t})\\
 &=\pnt[S]\times \pnt \langle\denote{s},\denote{t}\rangle & \text{inductive definition of }\pnt[S\times T]\\
 &=(\pnt[S]\denote{s},\pnt\denote{t})\\
 &=(\denote{v_1},\denote{v_2})
\end{align*}
for some values $v_1,v_2$. 
By the induction hypothesis it is therefore the case that $(s,N)\to v_1$ and $(t,N)\to v_2$ for any $N\in \N$ and it follows that $(s\otimes t, N)\to (v_1,v_2)$ by definition of the reduction relation $\to$.
\item $\lambda$-\textbf{abstraction.} If $\vdash \lambda x:\Code{S}.~t: \Code{S\to T}$, then the term $\lambda x:\Code{S}.~t$ is a value, and thus $(\lambda x:\Code{S}.~t,N)\to \lambda x:\Code{S}.~t$ trivially.
\item \textbf{Head.} Suppose that $\vdash \hd(t):\Code{T}$, and that $\pnt(\denote{\hd(t)})=\denote{v_1}$ for some value $v_1$.   Since the last applied rule has the premise $\vdash t:\stype \Code{T}$, and given the semantics of $\hd$, it must be the case that for any $N\geq 1$, $\left(\pnt[T\times R^+]\right)^N\circ \pi_{1:N}(t)=\denote{((v_1,w_1),\ldots, (v_N,w_N))}$ for some values $v_i,w_i$. By the induction hypothesis it must be the case that $(t,N)\to ((v_1,w_1),\ldots, (v_N,w_N))$, and thus that $(\hd(t),N)\to v_1$.
\item \textbf{Weight.} The proof is the same as that for \(\Code{hd}\). Suppose that $\vdash \wt(t):\Code{T}$, and that $\pnt(\denote{\wt(t)})=\denote{w_1}$ for some weight $w_1 \geq 0$.   Since the last applied rule has the  premise $\vdash t:\stype \Code{T}$, and given the semantics of $\wt$, it must be the case that for any $N\geq 1$, $\left(\pnt[T\times R^+]\right)^N\circ \pi_{1:N}(t)=\denote{((v_1,w_1),\ldots, (v_N,w_N))}$ for some values $v_i,w_i$. By the induction hypothesis it must be the case that $(t,N)\to ((v_1,w_1),\ldots, (v_N,w_N))$, and thus that $(\wt(t),N)\to w_1$.
\item \textbf{Pushforward.} Suppose that $\vdash \map(t,s):\stype\Code{T}$ and that $\pnt[\stype T](\denote{\map(t,s)})=\denote{((v_1,w_1),\ldots,(v_n,w_n)}$. The premises of the last applied rule must have been $\vdash s:\stype\Code{S}$ and $\vdash t: \Code{S\to T}$, and it follows from the semantics of $\map$ that $\denote{v_i}=\pnt\denote{t(\hd(\tl^{i-1}(s)))})$ and $\denote{w_i}=\denote{\wt(\tl^{i-1}(s))}$. It follows from the induction hypothesis that $(t(\hd(\tl^{i-1}(s)),N)\to v_i$ and $(wt(\tl^{i-1}(s)),N)\to w_i$, and thus by the definition of $\to$ we have that $(\map(t,s),N)\to ((v_1,w_1),\ldots,(v_n,w_n)$.
\end{enumerate}
\end{appendixproof}

\begin{appendixproof}[\cref{thm:sound_equivalence}]\label{proof:sound_equivalence}
\noindent \textbf{Standard rules}:
\begin{enumerate}
\item \textbf{\(\beta\)- and \(\eta\)-equivalence.}
\begin{gather*}
\denote{\Gamma \vdash (\lambda x : \Code{S} . t)(s) : \Code{T}} = \denote{\Gamma \vdash t[x\leftarrow s] : T}, \\
\denote{\Gamma \vdash \lambda x : \Code{S} . t(x) : \Code{S} \to \Code{T}} = \denote{\Gamma \vdash t : \Code{S} \to \Code{T}}
\end{gather*}
The soundness of \(\beta\)- and \(\eta\)-equivalence is well-known and immediate from the properties of exponential objects.

\item \textbf{\(\Code{let}\)-reduction.}
\begin{gather*}
\denote{\Gamma, s : \Code{S} \vdash \Code{let}\:x=s\:\Code{in}\:t : \Code{T}} = \denote{\Gamma, s : \Code{S} \vdash (\lambda x:\Code{S}.t)(s) : \Code{T}}
\end{gather*}
True by definition of the denotational semantics of \(\Code{let}\).

\item \textbf{Projections.}
\begin{gather*}
\denote{\Gamma \vdash \Code{fst}((s, t)) : \Code{S}} = \denote{\Gamma \vdash s : \Code{S}}, \\
\denote{\Gamma \vdash \Code{snd}((s, t)) : \Code{T}} = \denote{\Gamma \vdash t : \Code{T}}
\end{gather*}
Immediate from the properties of Cartesian products.
\end{enumerate}

\noindent \textbf{Congruence rules}:
\noindent Trivial in the denotational setting: if \(\denote{\Gamma \vdash s : \Code{S}} = \denote{\Gamma \vdash s' : \Code{S}}\) have identical semantics, then clearly, for any built-in operation \(\Code{op} : \Code{S} \to \Code{T}\), \(\denote{\Gamma \vdash \Code{op}(s) : \Code{T}} = \denote{\Gamma \vdash \Code{op}(s') : \Code{T}}\); the same extends to \(n\)-ary operations.

\noindent \textbf{Coinductive definitions}:
\begin{enumerate}
\item \textbf{Map.}
\begin{gather*}
\denote{\Gamma \vdash \Code{hd}(\Code{map}(s, t)) : \Code{T}} = \denote{\Gamma \vdash s(\Code{hd}(t)) : \Code{T}}, \\
\denote{\Gamma \vdash \Code{wt}(\Code{map}(s, t)) : \Code{R^+}} = \denote{\Gamma \vdash s(\Code{wt}(t)) : \Code{R^+}}, \\
\denote{\Gamma \vdash \Code{tl}(\Code{map}(s, t)) : \stype \Code{T}} = \denote{\Gamma \vdash \Code{map}(\Code{tl}(t)) : \stype \Code{T}}
\end{gather*}
Immediate from the coinductive definition of \(\Code{map}\).

\item \textbf{Product.}
\begin{gather*}
\denote{\Gamma \vdash (\Code{hd}(s), \Code{hd}(t)) : \Code{S} \times \Code{T}} = \denote{\Gamma \vdash \Code{hd}(s \otimes t) : \Code{S} \times \Code{T}}, \\
\denote{\Gamma \vdash \wt(s)*\wt(t) : \Code{R^+}} = \denote{\Gamma \vdash \Code{wt}(s \otimes t) : \Code{R^+}}, \\
\denote{\Gamma \vdash \Code{tl}(s) \otimes \Code{tl}(t) : \stype (\Code{S} \times \Code{T})} = \denote{\Gamma \vdash \Code{tl}(s \otimes t) : \stype (\Code{S} \times \Code{T})}
\end{gather*}
Immediate from the coinductive definition of \(\otimes\).

\item \textbf{Thinning.}
\begin{gather*}
\denote{\Gamma \vdash \Code{hd}(\Code{thin}(n, t)) : \Code{T}} = \denote{\Gamma \vdash \Code{hd}(t) : \Code{T}}, \\
\denote{\Gamma \vdash \Code{wt}(\Code{thin}(n, t)) : \Code{R^+}} = \denote{\Gamma \vdash \Code{wt}(t) : \Code{R^+}}, \\
\forall n \in \N, \denote{\Gamma \vdash \Code{tl}(\Code{thin}(n, t)) : \stype \Code{T}} = \denote{\Gamma \vdash \Code{thin}(n, \Code{tl}^n(t)) : \stype \Code{T}}, \\
\denote{\Gamma \vdash \Code{thin}(1, t) : \stype \Code{T}} = \denote{\Gamma \vdash t : \stype \Code{T}}
\end{gather*}
Immediate from the coinductive definition of \(\Code{thin}\).

\item \textbf{Pseudorandom number generators.}
\begin{gather*}
\denote{\Gamma \vdash \Code{hd}(\Code{prng}(s, t)) : \Code{T}} = \denote{\Gamma \vdash t : \Code{T}}, \\
\denote{\Gamma \vdash \Code{wt}(\Code{prng}(s, t)) : \Code{R^+}} = \denote{\Gamma \vdash 1 : \Code{R^+}}, \\
\denote{\Gamma \vdash \Code{tl}(\Code{prng}(s, t)) : \stype \Code{T}} = \denote{\Gamma \vdash \Code{prng}(s, s(t)) : \stype \Code{T}}
\end{gather*}
Immediate from the coinductive definition of \(\Code{prng}\).

\item \textbf{Reweighting.}
\begin{gather*}
\denote{\Gamma \vdash \Code{hd}(\Code{reweight}(s, t)) : \Code{T}} = \denote{\Gamma \vdash \Code{hd}(t) : \Code{T}}, \\
\denote{\Gamma \vdash \Code{wt}(\Code{reweight}(s, t)) : \Code{R^+}} = \denote{\Gamma \vdash s(\Code{hd}(t)) * \Code{wt}(t) : \Code{R^+}}, \\
\denote{\Gamma \vdash \Code{tl}(\Code{reweight}(s, t)) : \stype \Code{T}} = \denote{\Gamma \vdash \Code{reweight}(s, \Code{tl}(t)) : \stype \Code{T}}
\end{gather*}
Immediate from the coinductive definition of \(\Code{reweight}\).
\end{enumerate}

\textbf{Composition rules}:
\begin{enumerate}
\item \textbf{Thinning over thinning.}
\begin{gather*}
\denote{\Gamma \vdash \Code{thin}(n, \Code{thin}(m, t)) : \stype \Code{T}} = \denote{\Gamma \vdash \Code{thin}(n*m, t) : \stype \Code{T}}
\end{gather*}
For any possible value of the context \(\gamma \in \denote{\Gamma}\), we show equality between the elements \(\denote{\Gamma \vdash \Code{thin}(n, \Code{thin}(m, t)) : \stype \Code{T}}(\gamma)\) and \(\denote{\Gamma \vdash \Code{thin}(n*m, t) : \stype \Code{T}}(\gamma)\) of \(\denote{\stype T}\) coinductively. As all of the arguments we will make have precisely the same structure, we will only give that structure in full detail for this proof; for the rest, we will only present the bisimulation which gives our result.

We show this result by constructing, for each \(\gamma \in \denote{\Gamma}\), a bisimulation \(R(\gamma) \subseteq \denote{\stype T} \times \denote{\stype T}\). This is a set of samplers satisfying three properties:
\begin{enumerate}
\item \(\forall (s, t) \in R(\gamma), \pi_1(\mathrm{unfold}_{\Code{T}}(s)) = \pi_1(\mathrm{unfold}_{\Code{T}}(t))\); that is, the head of \(s\) and the head of \(t\) are the same
\item \(\forall (s, t) \in R(\gamma), \pi_2(\mathrm{unfold}_{\Code{T}}(s)) = \pi_2(\mathrm{unfold}_{\Code{T}}(t))\); that is, the first weight of \(s\) and the first weight of \(t\) are the same
\item \(\forall (s, t) \in R(\gamma), (\pi_3(\mathrm{unfold}_{\Code{T}}(s)), \pi_3(\mathrm{unfold}_{\Code{T}}(t))) \in R(\gamma)\); that is, applying \(\Code{tl}\) to two samplers in the bisimulation yields two more samplers in the bisimulation.
\end{enumerate}

The structure of this bisimulation \(R(\gamma)\) is typically found by applying \(\Code{tl}\) to both sides of the equivalence we wish to show, and then applying the rules we have previously shown (typically, the coinductive definitions of each operation, in this case \(\Code{thin}\)) to simplify what results. For example, in this case, we can simplify
\[
\denote{\Gamma \vdash \Code{tl}(\Code{thin}(n, \Code{thin}(m, t))) : \stype \Code{T}} = \denote{\Gamma \vdash \Code{thin}(n, \Code{thin}(m, \Code{tl}^{m*n}(t))) : \stype \Code{T}}
\]
and
\[
\denote{\Gamma \vdash \Code{tl}(\Code{thin}(n*m, t)) : \stype \Code{T}} = \denote{\Gamma \vdash \Code{thin}(n*m, \Code{tl}^{n*m}(t)) : \stype \Code{T}}.
\]
This suggests as a bisimulation the following set:
\[
R(\gamma) = \left\{\denote{(\Gamma \vdash \Code{thin}(n, \Code{thin}(m, \Code{tl}^k(t))) : \stype \Code{T}}(\gamma), \denote{\Gamma \vdash \Code{thin}(n*m, \Code{tl}^k(t)) : \stype \Code{T})}(\gamma) \mid k \in \N\right\}.
\]
In future, as this expression is quite crowded, we will drop the dependence on \(\gamma\).

We must now show that this is a valid bisimulation. First, we must show that applying \(\Code{hd}\) and \(\Code{wt}\) to each of these programs yields the same result, which is always immediate. In this case, applying the rule we had previously referred to as the coinductive definition of \(\Code{thin}\) gives
\[
\denote{\Gamma \vdash \Code{hd}(\Code{thin}(n, \Code{thin}(m, \Code{tl}^k(t)))) : \Code{T}} = \denote{\Gamma \vdash \Code{hd}(\Code{thin}(m, \Code{tl}^k(t))) : \Code{T}} = \denote{\Gamma \vdash \Code{hd}(\Code{tl}^k(t)) : \Code{T}}
\]
and
\[
\denote{\Gamma \vdash \Code{hd}(\Code{thin}(n*m, \Code{tl}^k(t))) : \Code{T}} = \denote{\Gamma \vdash \Code{hd}(\Code{tl}^k(t)) : \Code{T}};
\]
The same argument exactly applies for \(\Code{wt}\):
\[
\denote{\Gamma \vdash \Code{wt}(\Code{thin}(n, \Code{thin}(m, \Code{tl}^k(t)))) : \Code{R^+}} = \denote{\Gamma \vdash \Code{wt}(\Code{thin}(m, \Code{tl}^k(t))) : \Code{R^+}} = \denote{\Gamma \vdash \Code{wt}(\Code{tl}^k(t)) : \Code{R^+}}
\]
and
\[
\denote{\Gamma \vdash \Code{wt}(\Code{thin}(n*m, \Code{tl}^k(t))) : \Code{R^+}} = \denote{\Gamma \vdash \Code{wt}(\Code{tl}^k(t)) : \Code{R^+}};
\]

Finally, we must show that applying \(\Code{tl}\) to each of these expressions yields another element of the bisimulation. This is essentially the same argument as the one which led us to the bisimulation \(R\):
\begin{gather*}
\denote{\Gamma \vdash \Code{tl}(\Code{thin}(n, \Code{thin}(m, \Code{tl}^{k}(t)))) : \stype \Code{T}} \\
= \denote{\Gamma \vdash \Code{thin}(n, \Code{tl}^{n}(\Code{thin}(m, \Code{tl}^k(t)))) : \stype \Code{T}} \\
= \denote{\Gamma \vdash \Code{thin}(n, \Code{thin}(m, \Code{tl}^{n*m+k}(t))) : \stype \Code{T}}
\end{gather*}
and
\begin{gather*}
\denote{\Gamma \vdash \Code{tl}(\Code{thin}(n*m, \Code{tl}^k(t))) : \stype \Code{T}} = \\
\denote{\Gamma \vdash \Code{thin}(n*m, \Code{tl}^{n*m+k}(t)) : \stype \Code{T}}
\end{gather*}
Therefore, for any \(\gamma\), applying \(\Code{tl}\) will yield another element of our bisimulation, and so our proof is complete. Using this proof as a reference, we will abbreviate the remainder of the bisimulation proofs in the Appendix, as the structure of each argument is identical.

\item \textbf{Map over map.}
\[
\denote{\Gamma \vdash \Code{map}(g, \Code{map}(f, t)) : \stype \Code{T}} = \denote{\Gamma \vdash \Code{map}(g \circ f, t) : \stype \Code{T}}
\]
Applying the coinductive definition of \(\Code{map}\), we can easily see
\[
\denote{\Gamma \vdash \Code{tl}(\Code{map}(g, \Code{map}(f, t))) : \stype \Code{T}} = \denote{\Gamma \vdash \Code{map}(g, \Code{map}(f, \Code{tl}(t))) : \stype \Code{T}}
\]
and
\[
\denote{\Gamma \vdash \Code{tl}(\Code{map}(g \circ f, t)) : \stype \Code{T}} = \denote{\Gamma \vdash \Code{map}(g \circ f, \Code{tl}(t)) : \stype \Code{T}};
\]
which suggests the  bisimulation
\[
R = \left\{(\denote{\Gamma \vdash \Code{map}(g, \Code{map}(f, \Code{tl}^n(t))) : \stype \Code{T}}, \denote{\Gamma \vdash \Code{map}(g \circ f, \Code{tl}^n(t)) : \stype \Code{T})} \mid n \in \N\right\},
\]
This bisimulation is easily verified: simply apply \(\Code{hd}\) to both sides and reduce by applying the coinductive definition of \(\Code{map}\) and we will see that we obtain two equal expressions; apply \(\Code{wt}\) to both sides and reduce by applying the coinductive definition of \(\Code{map}\), and two equal expressions will result; and finally apply \(\Code{tl}\) to both sides and reduce by applying the coinductive definition of \(\Code{map}\), and we will see that the resulting pair is also included within this bisimulation.

\item \textbf{Reweighting over reweighting.}
\begin{gather*}
\denote{\Gamma \vdash \Code{reweight}(g, \Code{reweight}(f, t)) : \stype \Code{T}} = \denote{\Gamma \vdash \Code{reweight}(f \cdot g, t) : \stype \Code{T}}
\end{gather*}
Applying \(\Code{tl}\) to both sides and using the previous rule relating \(\Code{tl}\) and \(\Code{reweight}\), we easily obtain
\[
\denote{\Gamma \vdash \Code{tl}(\Code{reweight}(g, \Code{reweight}(f, t))) : \stype \Code{T}} = \denote{\Gamma \vdash \Code{reweight}(g, \Code{reweight}(f, \Code{tl}(t))) : \stype \Code{T}}
\]
and
\[
\denote{\Gamma \vdash \Code{tl}(\Code{reweight}(g \cdot f, t)) : \stype \Code{T}} = \denote{\Gamma \vdash \Code{reweight}(g \circ f, \Code{tl}(t)) : \stype \Code{T}}.
\]
Equivalence then follows from the bisimulation
\[
R = \left\{(\denote{\Gamma \vdash \Code{reweight}(g, \Code{reweight}(f, \Code{tl}^m(t))) : \stype \Code{T}}, \denote{\Gamma \vdash \Code{reweight}(g \circ f, \Code{tl}^m(t)) : \stype \Code{T})} \mid m \in \N\right\}
\]
which is easily verified, giving our desired equality.

\item \textbf{Thinning over pseudorandom number generators.}
\begin{gather*}
\forall n \in \N, \denote{\Gamma \vdash \Code{thin}(n, \Code{prng}(s, t)) : \stype \Code{T}} = \denote{\Gamma \vdash \Code{prng}(s^n, t) : \stype \Code{T}}
\end{gather*}
Applying \(\Code{tl}\) to both sides of each expression and simplifying using the coinductive definitions of \(\Code{thin}\) and \(\Code{prng}\), we obtain 
\[
\denote{\Gamma \vdash \Code{tl}(\Code{thin}(n, \Code{map}(s, t))) : \stype \Code{T}} = \denote{\Gamma \vdash \Code{thin}(n, \Code{map}(s, \Code{tl}^n(t))) : \stype \Code{T}}
\]
and
\[
\denote{\Gamma \vdash \Code{tl}(\Code{map}(s, \Code{thin}(n, t))) : \stype \Code{T}} = \denote{\Gamma \vdash \Code{map}(s, \Code{thin}(n, \Code{tl}^n(t))) : \stype \Code{T}}.
\]
This suggests the choice of bisimulation
\[
R = \left\{(\denote{\Gamma \vdash \Code{thin}(n, \Code{map}(s, \Code{tl}^m(t))) : \stype \Code{T}}, \denote{\Gamma \vdash \Code{map}(s, \Code{thin}(n, \Code{tl}^m(t))) : \stype \Code{T})}) \mid m \in \N\right\}
\]
which is easily verified and gives our result.

\item \textbf{Thinning over map.}
\begin{gather*}
\denote{\Code{\Gamma \vdash \Code{thin}(n, \Code{map}(s, t)) : \stype T}} = \denote{\Code{\Gamma \vdash \Code{map}(s, \Code{thin}(n, t)) : \stype T}}
\end{gather*}
Using the coinductive definitions of \(\Code{map}\) and \(\Code{thin}\), we obtain
\[
\denote{\Gamma \vdash \Code{tl}(\Code{thin}(n, \Code{map}(s, t)) : \stype \Code{T}} = \denote{\Gamma \vdash \Code{thin}(n, \Code{map}(s, \Code{tl}^n(t))) : \stype \Code{T}}
\]
and
\[
\denote{\Gamma \vdash \Code{tl}(\Code{map}(s, \Code{thin}(n, t))) : \stype \Code{T}} = \denote{\Gamma \vdash \Code{map}(s, \Code{thin}(n, \Code{tl}^n(t))) : \stype \Code{T}}.
\]
The desired result follows from
\[
R = \{(\denote{\Gamma \vdash \Code{thin}(n, \Code{map}(s, \Code{tl}^m(t))) : \stype \Code{T}}, \denote{\Gamma \vdash \Code{map}(s, \Code{thin}(n, \Code{tl}^m(t))) : \stype \Code{T}}) \mid m \in \N\}
\]
which is easily seen to be a valid bisimulation.
\end{enumerate}

\textbf{Product rules}:
\begin{enumerate}

\item \textbf{Thinning.}
\begin{gather*}
\denote{\Gamma \vdash \Code{thin}(n, s) \otimes \Code{thin}(n, t) : \stype(\Code{S} \times \Code{T})}= \denote{\Gamma \vdash \Code{thin}(n, s \otimes t) : \stype (\Code{S} \times \Code{T})}
\end{gather*}
Use the coinductive definition of \(\Code{thin}\) to show
\[
\denote{\Gamma \vdash \Code{tl}(\Code{thin}(n, s) \otimes \Code{thin}(n,t)) : \stype(\Code{S} \times \Code{T})} = \denote{\Gamma \vdash  \Code{thin}(n, \Code{tl}^n(s)) \otimes \Code{thin}(n, \Code{tl}^n(t)) : \stype (\Code{S} \times \Code{T})}
\]
and
\[
\denote{\Gamma \vdash \Code{tl}(\Code{thin}(n, s \otimes t)) : \stype(\Code{S} \times \Code{T})} = \denote{\Gamma \vdash \Code{thin}(n, \Code{tl}^n(s) \otimes \Code{tl}^n(t)) : \stype(\Code{S} \times \Code{T})},
\]
which suggests
\begin{align*}
R = \{& (\denote{\Gamma \vdash \Code{thin}(n, \Code{tl}^m(s)) \otimes \Code{thin}(n, \Code{tl}^m(t)) : \stype (\Code{S} \times \Code{T})}, \\
& \denote{\Gamma \vdash \Code{thin}(n, \Code{tl}^m(s \otimes t)) : \stype(\Code{S} \times \Code{T})}) : m \in \N\}
\end{align*}
as a bisimulation.

\item \textbf{Map.}
\begin{gather*}
\denote{\Gamma \vdash s \otimes \Code{map}(g, t') : \stype(\Code{S} \times \Code{T})} = \denote{\Gamma \vdash \Code{map}(id_{\Code{S}} \times g, s \otimes t') : \stype (\Code{S} \times \Code{T})}
\end{gather*}
Applying \(\Code{tl}\) to each expression yields
\[
\denote{\Gamma \vdash \Code{tl}(s \otimes \Code{map}(g, t')) : \stype (\Code{S} \times \Code{T})} = \denote{\Gamma \vdash \Code{tl}(s) \otimes \Code{map}(g, \Code{tl}(t')) : \stype (\Code{S} \times \Code{T})}
\]
and
\[
\denote{\Gamma \vdash \Code{tl}(\Code{map}(id_{\Code{S}} \times g, s \otimes t')) : \stype (\Code{S} \times \Code{T}} = \denote{\Gamma \vdash \Code{map}(id_{\Code{S}} \times g, \Code{tl}(s) \otimes \Code{tl}(t')) : \stype (\Code{S} \times \Code{T})},
\]
suggesting
\begin{align*}
R = \{ & (\denote{\Gamma \vdash \Code{tl}^m(s) \otimes \Code{map}(g, \Code{tl}^m(s')) : \stype (\Code{S} \times \Code{T})}, \\
& \denote{\Gamma \vdash \Code{map}(id_{S'} \times g, \Code{tl}^m(s), \Code{tl}^m(s))) : \stype (\Code{S} \times \Code{T})}) \mid m \in \N\}
\end{align*}
as a bisimulation.

\begin{gather*}
\denote{\Gamma \vdash \Code{map}(f, t) \otimes s' : \stype (\Code{S} \times \Code{T})} = \denote{\Gamma \vdash \Code{map}(f \times id_T, t \otimes s') : \stype (\Code{S} \times \Code{T})}
\end{gather*}
Same proof as previous.

\item \textbf{Reweighting.}
\begin{gather*}
\denote{\Gamma \vdash s \otimes \Code{reweight}(g, t') : \stype (\Code{S} \times \Code{T})} = \denote{\Gamma \vdash \Code{reweight}(1_S \cdot g, s \otimes t') : \stype (\Code{S} \times \Code{T})}
\end{gather*}
Applying \(\Code{tl}\) to both sides and simplifying using the coinductive definition of \(\Code{reweight}\) gives 
\[
\denote{\Gamma \vdash \Code{tl}(s \otimes \Code{reweight}(g, t')) : \stype (\Code{S} \times \Code{T})} = \denote{\Gamma \vdash \Code{tl}(s) \otimes \Code{reweight}(g, \Code{tl}(t')) : \stype (\Code{S} \times \Code{T})}
\]
and
\[
\denote{\Gamma \vdash \Code{tl}(\Code{reweight}(1_{\Code{S}} \cdot g, s \otimes t')) : \stype(\Code{S} \times \Code{T})} = \denote{\Gamma \vdash \Code{reweight}(1_{\Code{S}} \cdot g, \Code{tl}(s) \otimes \Code{tl}(t')) : \stype(\Code{S} \times \Code{T})}.
\]
Choosing the bisimulation
\begin{align*}
R = \{ & (\denote{\Gamma \vdash \Code{tl}^m(s) \otimes \Code{reweight}(g, \Code{tl}^m(t')) : \stype (\Code{S} \times \Code{T})}, \\
& \denote{\Gamma \vdash \Code{reweight}(1_{\Code{S}} \cdot g, \Code{tl}^m(s) \otimes \Code{tl}^m(t')) : \stype (\Code{S} \times \Code{T})}) \mid m \in \N\},
\end{align*}
our result follows.

\begin{gather*}
\denote{\Gamma \vdash \Code{reweight}(f, t) \otimes s' : \stype(\Code{S} \times \Code{T})} = \denote{\Gamma \vdash \Code{reweight}(f \cdot 1_{\Code{T}}, s' \otimes t) : \stype (\Code{S} \times \Code{T})}
\end{gather*}
Same proof as previous.

\item \textbf{Pseudorandom number generators.}
\begin{gather*}
\denote{\Gamma \vdash \Code{prng}(f, a) \otimes \Code{prng}(g, b) : \stype(\Code{S} \times \Code{T})} = \denote{\Gamma \vdash \Code{prng}(f \times g, (a, b)) : \stype (\Code{S} \times \Code{T})}
\end{gather*}
Using the coinductive definition of \(\Code{prng}\), we quickly obtain
\[
\denote{\Gamma \vdash \Code{tl}(\Code{prng}(f, a) \otimes \Code{prng}(g, b)) : \stype(\Code{S} \times \Code{T})} = \denote{\Gamma \vdash \Code{prng}(f, f(a)) \otimes \Code{prng}(g, g(b)) : \stype(\Code{S} \times \Code{T})}
\]
and
\[
\denote{\Gamma \vdash \Code{tl}(\Code{prng}(f \times g, (a, b))) : \stype (\Code{S} \times \Code{T})} = \denote{\Gamma \vdash \Code{prng}(f \times g, (f \times g)(a, b)) : \stype(\Code{S} \times \Code{T})},
\]
suggesting the bisimulation
\begin{align*}
R = \{ & (\denote{\Gamma \vdash \Code{prng}(f, f^m(a)) \otimes \Code{prng}(g, g^m(b)) : \stype(\Code{S} \times \Code{T})}, \\
& \denote{\Gamma \vdash \Code{prng}(f \times g, (f \times g)^m(a,b)) : \stype (\Code{S} \times \Code{T})}) \mid m \in \N\}
\end{align*}
which gives our desired result.
\end{enumerate}
\end{appendixproof}

\begin{appendixproof}[\cref{prop:nestedselfproduct}]\label{proof:nestedselfproduct}
Expanding \cref{eq:selfproduct}, for any well-typed sampler \(\Gamma \vdash s : \stype \Code{S}\), the nested self-product \((s^m)^n\) is defined as
\[
\Code{thin}(n, \Code{thin}(m, s \otimes \Code{tl}(s) \otimes \dots \otimes \Code{tl}^{m-1}(s)) \otimes \dots \otimes \Code{tl}^{n-1}(\Code{thin}(m, s \otimes \Code{tl}(s) \otimes \dots \otimes \Code{tl}^{m-1}(s)))).
\]
Applying the rule \(\Gamma \vdash \Code{tl}(\Code{thin}(m, t)) \approx \Code{thin}(m, \Code{tl}^{m}(t)) : \stype \Code{T}\) from \cref{table:equivalencerules} on the innermost expressions, it follows that this program is equivalent in the context \(\Gamma\) to
\[
\Code{thin}(n, \Code{thin}(m, \Code{tl}^0(s) \otimes \dots \otimes \Code{tl}^{m-1}(s)) \otimes \dots \otimes \Code{thin}(m, \Code{tl}^{mn-m}(s) \otimes \dots \otimes \Code{tl}^{mn-1}(s))).
\]
Next, applying the rule \(\Gamma \vdash \Code{thin}(m, s \otimes t) \approx \Code{thin}(m, s) \otimes \Code{thin}(m, t) : \stype (\Code{S} \times \Code{T})\), we see that the nested self-product is equivalent to
\[
\Code{thin}(n, \Code{thin}(m, \Code{tl}^0(s) \otimes \Code{tl}^1(s) \otimes \dots \otimes \Code{tl}^{mn-1}(s))).
\]
Applying the rule \(\Gamma \vdash \Code{thin}(n, \Code{thin}(m, t)) \approx \Code{thin}(mn, t) : \stype \Code{T}\) for composition of \(\Code{thin}\) yields
\[
\Code{thin}(mn, \Code{tl}^0(s) \otimes \Code{tl}^1(s) \otimes \dots \otimes \Code{tl}^{mn-1}(s)),
\]
and the above is precisely the definition of the self-product \(s^{mn}\).
\end{appendixproof}

\begin{appendixproof}[\cref{prop:mapreweightselfproduct}]\label{proof:mapreweightselfproduct}
\begin{itemize}
\item \textbf{Map}: Applying the definition of the self-product \cref{eq:selfproduct}, the syntax \(\Code{map}(f, s)^n\) is shorthand for the sampler
\[
\Code{thin}(n, \Code{map}(f, s) \otimes \Code{tl}(\Code{map}(f, s)) \otimes \dots \otimes \Code{tl}^{n-1}(\Code{map}(f, s)))
\]
In a context \(\Gamma\) in which this sampler is well-typed, applying the rule \(\Gamma \vdash \Code{tl}(\Code{map}(f, s)) \approx \Code{map}(f, \Code{tl}(s)) : \stype \Code{T}\) shows that the above sampler is equivalent to
\[
\Code{thin}(n, \Code{map}(f, \Code{tl}^0(s)) \otimes \dots \otimes \Code{map}(f, \Code{tl}^{n-1}(s))).
\]
For the purposes of this proof, abbreviate the \(n\)-fold Cartesian product of a program \(f : \Code{S} \to \Code{T}\) as \(f^{\times n} : \Code{S}^n \to \Code{T}^n\). Applying the rule \(\Gamma \vdash \Code{map}(f, s) \otimes \Code{map}(g, t) \approx \Code{map}(f \times g, (s, t)) : \stype(\Code{S} \times \Code{T})\), this sampler can also be written in the equivalent form
\[
\Code{thin}(n, \Code{map}(f^{\times n}, \Code{tl}^0(s) \otimes \dots \otimes \Code{tl}^{n-1}(s))).
\]
Finally, applying the rule \(\Gamma \vdash \Code{thin}(n, \Code{map}(f, s)) \approx \Code{map}(f, \Code{thin}(n, s)) : \stype \Code{S}\) yields
\[
\Code{map}(f^{\times n}, \Code{thin}(n, \Code{tl}^0(s) \otimes \dots \otimes \Code{tl}^{n-1}(s))))
\]
which is, by the definition of the self-product, our desired result \(\Code{map}(f^{\times n}, s^n)\).

\item \textbf{Reweight}: This proof proceeds the same as the above, but with \(\Code{map}\) replaced with \(\Code{reweight}\) and the Cartesian product \(\times\) replaced with the pointwise product \(\cdot\); nevertheless, we will go through it. Applying the definition of the self-product \cref{eq:selfproduct}, the syntax \(\Code{reweight}(f, s)^n\) is shorthand for
\[
\Code{thin}(n, \Code{reweight}(f, s) \otimes \Code{tl}(\Code{reweight}(f, s)) \otimes \dots \otimes \Code{tl}^{n-1}(\Code{reweight}(f, s)))
\]
In a context \(\Gamma\) in which this sampler is well-typed, applying the rule \(\Gamma \vdash \Code{tl}(\Code{reweight}(f, s)) \approx \Code{reweight}(f, \Code{tl}(s)) : \stype \Code{T}\) shows that the above sampler is equivalent to
\[
\Code{thin}(n, \Code{reweight}(f, \Code{tl}^0(s)) \otimes \dots \otimes \Code{reweight}(f, \Code{tl}^{n-1}(s))).
\]
For the purposes of this proof, abbreviate the \(n\)-fold pointwise product of a program \(f : \Code{S} \to \Code{R}\) as \(f^{\cdot n} : \Code{S}^n \to \Code{R}\). Applying the rule \(\Gamma \vdash \Code{reweight}(f, s) \otimes \Code{reweight}(g, t) \approx \Code{reweight}(f \cdot g, (s, t)) : \stype(\Code{S} \times \Code{T})\), this sampler can also be written in the equivalent form
\[
\Code{thin}(n, \Code{reweight}(f^{\cdot n}, \Code{tl}^0(s) \otimes \dots \otimes \Code{tl}^{n-1}(s))).
\]
Finally, applying the rule \(\Gamma \vdash \Code{thin}(n, \Code{reweight}(f, s)) \approx \Code{reweight}(f, \Code{thin}(n, s)) : \stype \Code{S}\) yields
\[
\Code{reweight}(f^{\cdot n}, \Code{thin}(n, \Code{tl}^0(s) \otimes \dots \otimes \Code{tl}^{n-1}(s))))
\]
which is, by the definition of the self-product, our desired result \(\Code{reweight}(f^{\cdot n}, s^n)\).
\end{itemize}
\end{appendixproof}

\begin{appendixproof}[\cref{prop:continuous_case}]\label{proof:continuous_case}
Let $g: Y\to \R$ be a bounded continuous function. Then $g\circ f: X\to \R$ is also bounded continuous, and it follows from the definition of weak convergence and of $\varepsilon$ that
\begin{align*}
\lim_{n\to\infty} \int_X g ~d\widehat{f\circ \sigma}_n&=\lim_{n\to\infty}\int_X g\circ f ~d\hat{\sigma}_n &\text{by definition}\\
&= \int_X g\circ f~d\mu&\text{since }\varepsilon_X(\sigma)=\mu\\
&=\int_X g~d f_\ast\mu&\text{change of variable}
\end{align*}
Thus $\widehat{f\circ\sigma}_n \longrightarrow f_\ast\mu$ weakly,  \ie $\varepsilon(f\circ\sigma)=f_\ast(\mu)$.
\end{appendixproof}

\begin{appendixproof}[\cref{thm:targeting_sound}]\label{proof:targeting_sound}
By induction on the derivation.  
\begin{enumerate}[(i)]
\item \textbf{Built-in samplers.} These axioms are true by assumption.
\item \textbf{Equivalence.} The fact that equivalent terms target the same measure is a simple consequence of the definition of targeting and of \cref{thm:sound_equivalence}.
\item \textbf{Tail.} The fact that the tail of a sampler $\sigma$ targets the same measure as $\sigma$ is a simple consequence of the definition of targeting in terms of a limit.
\item \textbf{Pushforward.} If $\denote{f}$ is continuous for the standard topologies of the type system, the the rule is a direct consequence of \cref{prop:continuous_case}. If $\denote{f}$ is not continuous for the standard topologies of the type system, then either: (a) a measure $\mu(\gamma)$ assigns some mass to the boundary of an element of the partition making $\denote{f}$ piecewise continuous, in which case $\varepsilon{\denote{s}(\gamma)}$ will not converge to $\mu(\gamma)$ and the premise of the rule does not hold,  or (b) no measure $\mu(\gamma)$ assigns any mass to the boundary of an element of the partition making $\denote{f}$ continuous, in which case $\varepsilon{\denote{s}(\gamma)}=\mu(\gamma)$,  and the conclusion is again a consequence of \cref{prop:continuous_case}.
\item \textbf{Reweight.} This rule simply encodes the validity of importance sampling. Dropping the dependency in $\gamma$ for clarity of notation, and letting \(\nu = f \cdot \mu\) be the reweighted measure, the premise and side-condition of the rule together say that there exists $\alpha\in\R^+$ such that $f=\alpha \frac{d\mu}{d\nu}$, and that $f$ is bounded on the support of \(\mu\).  Since $\mu$ is a probability distribution, we have
\[
\int f~d\mu =\alpha\int   \frac{d\mu}{d\nu}~d\nu=\alpha \int ~d\nu=\alpha
\]
Moreover, since $s$ targets $\mu$ and $f$ is bounded continuous, we get by writing $x_i\defeq \pi_1(\pi_i(\denote{s})$ and $w_i\defeq \pi_2(\pi_i(\denote{s})$ that
\begin{align}
\alpha=\int f~d\mu=\lim_{N\to\infty}\frac{1}{N} \frac{\sum_{i=1}^Nf(x_i)w_i}{\sum_{k=1}^Nw_k}\label{eq:importance}
\end{align}
Letting $g$ be any bounded continuous function $\denote{\Code{S}}\to\R$ and noting that the pointwise product \(g.f\) is bounded continuous on the support of \(\mu\) and \(\nu\), we have
\begin{align*}
\int g~d\nu &= \frac{1}{\alpha}\int g.f~d\mu \\
&=\frac{1}{\alpha} \lim_{N\to\infty}\frac{1}{N}\sum_{i=1}^N g(x_i)f(x_i)\frac{w_i}{\sum_{k=1}^Nw_k} & \text{Since}~s~\text{targets}~\mu \\
&=\left(\lim_{N\to\infty}\frac{1}{N}\frac{\sum_{i=1}^N f(x_i) w_i}{\sum_{k=1}^Nw_k}\right)\inv\left( \lim_{N\to\infty}\frac{1}{N}\frac{\sum_{i=1}^N g(x_i)f(x_i)w_i}{\sum_{k=1}^Nw_k}\right) & \text{By}~\eqref{eq:importance} \\
&=\lim_{N\to\infty} \frac{1}{N}\frac{\sum_{i=1}^N g(x_i)f(x_i)w_i}{\sum_{i=1}^N f(x_i)w_i} \\
&\defeq \lim_{N\to\infty}\int g~d\widehat{\mathtt{reweight}(f,s)_n}
\end{align*}
In other words, $\mathtt{reweight}(f,s)$ targets $\nu$.
\item \textbf{Pseudorandom number generators.} The \texttt{prng} rule is just a restatement of the well-known ergodic theorem \cite[Theorem 9.6]{kallenberg1997foundations}.
\end{enumerate}
\end{appendixproof}

\newcommand{\Meas}{\mathbf{Meas}}
\newcommand{\Borel}{\mathsf{Borel}}
\begin{appendixproof}[\cref{prop:targeting_subtypes}]\label{proof:targeting_subtypes}
The first part of the proof follows immediately if we can show that $\Code{S}\sub\Code{T}$ implies that $\denote{\Code{S}}$ and $\denote{\Code{T}}$ are the same measurable space; this will be shown by induction on the sub-typing derivation. We start by showing that the functor $\Borel: \Top\to \Meas$ commutes with coproducts. This will prove the base case, the coproduct rule, and the last two rules of \cref{lang:subtyping}.

Let $X,Y$ be two topological spaces (we will use the same name for topological (\resp measurable) spaces and their topologies (\resp $\sigma$-algebras)). We use the $\pi$-$\lambda$ lemma to prove $\Borel(X+Y)=\Borel(X) + \Borel(Y)$.  First note that $\Borel(X+Y)=\sigma(X+Y)$ by definition. Since $X+Y$ is a topology,  it is trivially also a $\pi$-system,  and since $\Borel(X) + \Borel(Y)$ is a $\sigma$-algebra it is also trivially a $\lambda$-system. By definition, every open set $U$ in $X+Y$ has the property that $U=X\cap U$ is open in $X$, and is thus an element of $\Borel(X)$. Similarly $Y\cap U$ is open in $Y$ and thus belongs to $\Borel(Y)$. It follows that $U=(U\cap X)\uplus (U\cap Y)$ belongs to $\Borel(X) + \Borel(Y)$ by definition of the coproduct in $\Meas$.  The inclusion $\Borel(X+Y)\subseteq \Borel(X) + \Borel(Y)$ now follows from the $\pi$-$\lambda$ lemma.

Conversely,  every measurable $A$ in $\Borel(X) + \Borel(Y)$ is, by definition,  of the shape $(A\cap X)\uplus (A\cap Y)$ with $(A\cap X)\in\Borel(X)$ and $(A\cap Y)\in\Borel(Y)$. Using the $\pi$-$\lambda$ lemma it is easy to show that $\Borel(X)\subseteq \Borel(X+Y)$ and $\Borel(Y)\subseteq \Borel(X+Y)$, and it thus follows, since  $\Borel(X+Y)$ is closed under unions, that $A=(A\cap X)\uplus (A\cap Y)\in \Borel(X+Y)$ which proves $\Borel(X+Y)\supseteq \Borel(X) + \Borel(Y)$.

To show that the functor $\Borel: \Top\to \Meas$ commutes with products we need the extra assumption that the spaces are second-countable. A proof can then be found in \eg \cite[p244]{billingsley2013convergence}.  

For the second part of the proof, let $U$ be in the topology of $\denote{\Code{S}}$ but not in the topology of $\denote{\Code{T}}$. This means that $\partial_{\Code{T}}(U)=U\cap \mathrm{int}_{\Code{T}}(U)\neq \empty$ is open in $\denote{\Code{S}}$ (since it's the intersection of two open sets in $\denote{\Code{S}}$). In particular it is a continuity set in $\denote{\Code{S}}$ (since it is open, its interior is the empty set and it can therefore not have any $\mu$-mass).  By the Portmanteau lemma (which applies since the spaces are assumed to be metrizable) we must thus have
\[
\lim_{n\to\infty} \widehat{\denote{s}}_n\left(\partial_{\Code{T}}(U)\right) = \mu(\partial_{\Code{T}}(U))>0
\]
In particular, this is clearly impossible if $\denote{s}$ only visits $\partial_{\Code{T}}(U)$ finitely many times.
\end{appendixproof}

\fi

\end{document}